\documentclass[twocolumn,showpacs,amssymb,prd,superscriptaddress,nofootinbib]{revtex4-1}

\usepackage{graphicx,epsf, epsfig, amssymb}
\usepackage{bm}
\usepackage{longtable}
\usepackage{color}
\usepackage{hyperref}
\usepackage{amsfonts,amsmath,amssymb,mathrsfs}

\def\be{\begin{equation}}
\def\ee{\end{equation}}
\def\beq{\begin{eqnarray}}
\def\eeq{\end{eqnarray}}
\def\f{\frac}
\newcommand{\nn}{\nonumber}

\newcommand{\bgaln}{\begin{align}}
\newcommand{\bgeq}{\begin{equation}}

\newcommand{\Rmnum}[1]{\expandafter\@slowromancap\romannumeral #1@}

\begin{document}

\title {Gravitational radiation from compact binary systems in the massive Brans-Dicke theory of gravity}

\author{Justin Alsing} \email{justin.alsing@seh.ox.ac.uk}
\affiliation{Department of Physics, University of Oxford, Keble Road, Oxford OX1 3RH,
UK}

\author{Emanuele Berti}
\email{berti@phy.olemiss.edu}                                
\affiliation{Department of Physics and Astronomy, The University of
  Mississippi, University, MS 38677, USA}
\affiliation{California Institute of Technology, Pasadena, CA 91109,
  USA}

\author{Clifford M. Will}
\email{cmw@wuphys.wustl.edu}
\affiliation{McDonnell Center for the Space Sciences, Department of
  Physics, Washington University, St. Louis, MO 63130, USA}

\author{Helmut Zaglauer}
\affiliation{Astrium GmbH, 88039 Friedrichshafen, Germany}

\date{\today}

\begin{abstract}
  We derive the equations of motion, the periastron shift, and the
  gravitational radiation damping for quasicircular compact binaries
  in a massive variant of the Brans-Dicke theory of gravity. We also
  study the Shapiro time delay and the Nordtvedt effect in this
  theory. By comparing with recent observational data, we put bounds
  on the two parameters of the theory: the Brans-Dicke coupling
  parameter $\omega_{\rm BD}$ and the scalar mass $m_s$. We find that
  the most stringent bounds come from Cassini measurements of the
  Shapiro time delay in the Solar System, that yield a lower bound
  $\omega_{\rm BD}>40000$ for scalar masses
  $m_s<2.5\times10^{-20}\mathrm{eV}$ (or Compton wavelengths
  $\lambda_s=h/(m_sc)>5\times 10^{10}$~km), to 95\% confidence. In
  comparison, observations of the Nordtvedt effect using Lunar Laser
  Ranging (LLR) experiments yield $\omega_{\rm BD}>1000$ for
  $m_s<2.5\times10^{-20}\mathrm{eV}$. Observations of the orbital
  period derivative of the quasicircular white dwarf-neutron star
  binary PSR J1012+5307 yield $\omega_{\rm BD}>1250$ for
  $m_s<10^{-20}\mathrm{eV}$ ($\lambda_s>1.2\times 10^{11}$~km). A
  first estimate suggests that bounds comparable to the Shapiro time
  delay may come from observations of radiation damping in the
  eccentric white dwarf-neutron star binary PSR J1141-6545, but a
  quantitative prediction requires the extension of our work to
  eccentric orbits.
\end{abstract}
\maketitle

General relativity (GR) occupies a well earned place next to the
standard model as one of the two pillars of modern physics.  All
observational evidence to date supports GR as the correct classical
theory of gravitation, but there are countless attempts at developing
alternative theories of gravity. Two of the main motivations for these
efforts are the desire to formulate a fully quantizable theory of
gravity, and the quest to uncover the mechanisms underlying the dark
energy problem in cosmology. In addition, the vast majority of tests
of GR that have been carried out to date are in the weak-field, low
energy regime, but it is widely believed that GR may indeed break down
at higher energies. The direct observation of gravitational waves with
Earth- and space-based detectors will mark the dawn of a new era,
allowing us to probe gravity in the dynamical, strong-field
regime. For these reasons, the study of gravitational radiation in
modified theories of gravity has become a central issue.

One of the most popular and simple alternative theories of gravity is
scalar-tensor theory, in which gravity is mediated by both a scalar
and a tensor field, coupled together in a nontrivial manner through
the presence of a nonminimal coupling term in the action
\cite{Fujii:2003pa,EspositoFarese:2009ta,Clifton:2011jh}.  The
existence of scalar partners to the graviton is predicted in all
extra-dimensional theories, and scalar fields play a crucial role in
modern cosmology.  Scalar-tensor theories are consistent, have a
well-posed Cauchy problem, and respect many of the symmetries of GR.
They are also conformally equivalent to GR (if the coupling with
matter is nonstandard), allowing us to employ the same techniques used
to solve the Einstein field equations as long as we work in the
Einstein frame \cite{Fujii:2003pa,Clifton:2011jh}. Finally, generic
scalar-tensor theories can be shown to be equivalent to $f(R)$
theories \cite{Sotiriou:2008rp,DeFelice:2010aj}.  A good account of
the motivations behind scalar-tensor theories, including their
historical development, can be found in
\cite{Fujii:2003pa,Clifton:2011jh}.

String theory suggests the existence of massive but light scalar
fields (``axions'') with masses possibly as small as the Hubble scale
($\sim 10^{-33}$~eV). If we do indeed live in a ``string axiverse'',
CMB observations, galaxy surveys and measurements of black hole spins
may offer exciting experimental opportunities to set constraints on
the mass of these scalars \cite{Arvanitaki:2010sy,Kodama:2011zc}.

Here we are interested in the possibility of constraining the mass and
coupling of massive scalars via present (electromagnetic) and future
(gravitational-wave) observations of compact binaries.  Until
recently, calculations of gravitational radiation damping in
scalar-tensor theories (see
e.g.~\cite{Wagoner:1970vr,Will:1989sk,Brunetti:1998cc,Damour:1998jk})
have focused mostly on the {\em massless} case. Due to the interest of
light scalars in cosmology and high-energy physics, this restriction
has been dropped in more recent work.
For example it has been shown that resonant, superradiant effects
induced by light, massive scalars may produce ``floating orbits'' when
small compact objects inspiral into rotating black holes, leaving a
distinct signature in gravitational waves
\cite{Cardoso:2011xi,Yunes:2011aa}.

A commonly held belief is that only {\it mixed binaries} (i.e.,
binaries whose members have different gravitational binding energy)
can produce significant amounts of scalar gravitational
radiation. There are two reasons for this. The first is that, under
standard assumptions, dipole radiation is produced due to violations
of the strong equivalence principle when the binary members have
unequal ``sensitivities'': $s_1\neq s_2$. These sensitivities are
defined in Eq.~(\ref{sensitivity}) below, and they are related to the
gravitational binding energy of each binary member. In other words,
dipole radiation is produced when the system's center of mass is
offset with respect to the center of inertia (see
e.g. \cite{Clifton:2011jh}), so that mixed binaries and eccentric
binaries would be the best target to constrain scalar-tensor
theories. The second reason is the black hole no-hair theorem,
i.e. the fact that black hole solutions in scalar-tensor theories are
the same as in GR (see \cite{Sotiriou:2011dz} and references
therein). Building on earlier work by Jacobson \cite{Jacobson:1999vr},
Horbatsch and Burgess recently pointed out that slowly varying scalar
fields may violate the no-hair theorem, so that even black hole-black
hole binaries may produce dipole radiation
\cite{Horbatsch:2011ye}. They also developed a formalism to test
generic scalar-tensor theories using binary pulsars
\cite{Horbatsch:2011nh}.

For all these reasons, a study of gravitational radiation in massive
scalar-tensor theories is quite timely. In this paper we derive the
period derivative due to scalar and tensor radiation in theories with
a massive scalar field.  For simplicity we focus on circular binaries,
but (as we will see below) the generalization of our results to
eccentric binaries would be of great observational
interest\footnote{We will be working in units $\hbar=c=G=1$ throughout
  the paper. Greek indices will span both spatial and time components
  ${0,1,2,3}$.  Roman indices run over the spatial components
  ${1,2,3}$ only. We will adopt the metric signature $(-,+,+,+)$.}.

For the reader's convenience, here we give an executive summary of our
main results.  Consider a compact binary in circular orbit with
component masses $m_i$ and sensitivities $s_i$ $(i=1\,,2)$. Then the
period derivative due to the emission of scalar and tensor
gravitational waves in the massive Brans-Dicke theory is
\begin{align}
\label{period_decay_mbd}
\f{\dot{P}}{P}=-\f{8}{5}\f{\mu m^2}{r^4}\kappa_1-\f{\mu m}{r^3}\kappa_D\mathcal{S}^2\,,
\end{align}
where
\begin{align}
& \kappa_1=\mathcal{G}^2\left[12-6\xi+\xi\Gamma^2\left(\f{4\omega^2-m_s^2}{4\omega^2}\right)^2\Theta(2\omega-m_s)\right]\,, \nonumber \\
& \kappa_D=2\mathcal{G}\xi\f{\omega^2-m_s^2}{\omega^2}\Theta(\omega-m_s)\,,
\end{align}
$\Theta$ is the Heaviside function, $r$ is the separation of the
binary members, $m_s$ is the mass of the scalar field, $m=m_1+m_2$ and
$\mu=m_1 m_2/m$ are the total and reduced masses of the system,
$\mathcal{S}\equiv s_2-s_1$ and furthermore
\begin{align}
&\xi= \f{1}{2+\omega_{\rm BD}}\,, \nonumber \\
&\mathcal{G}= 1-\xi\left(s_1+s_2-2s_1s_2\right)\,, \nonumber \\
&\Gamma= 1-2\f{s_1m_2+m_1s_2}{m}\,. \nonumber
\end{align}
Note that scalar dipole radiation is emitted only when the binary's
orbital frequency $\omega>m_s$ and the difference in sensitivities
${\cal S}\neq 0$, while scalar quadrupole/monopole radiation is
emitted only when $2\omega>m_s$ and it also vanishes for two black
holes (since in that case $s_1=s_2=1/2$ and $\Gamma=0$). 
This result is only strictly valid in the limit of a very massive
($m_sr\gg1$) or very light ($m_sr\ll1$) scalar. However corrections
due to an intermediate mass scalar always enter with at least a factor
of the small parameter $\xi$, so this should be a relatively good
approximation for the full range of scalar masses.

\begin{figure*}[htb]
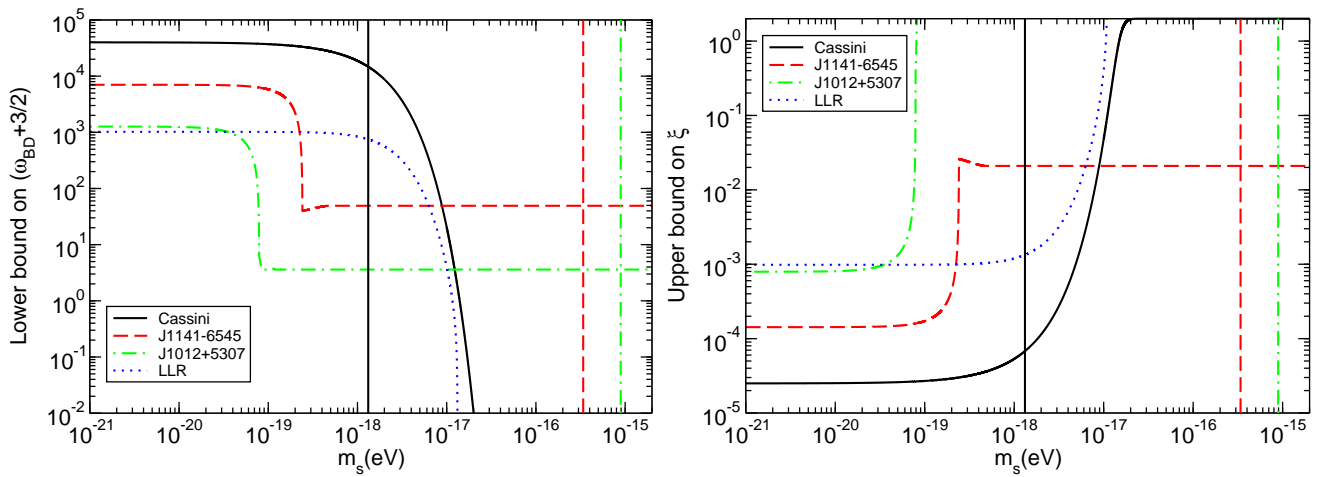

\begin{tabular}{cc}
\includegraphics[width=8.6cm,clip=true]{fig1a.eps} &
\includegraphics[width=8.6cm,clip=true]{fig1b.eps}
\end{tabular}
\caption{Left: Lower bound on $(\omega_{\rm BD}+3/2)$ as a function of
  the mass of the scalar $m_s$ from the Cassini mission data (black
  solid line; cf. \cite{Perivolaropoulos:2009ak}), period derivative
  observations of PSR J1141-6545 (dashed red line) and PSR J1012+5307
  (dot-dashed green line), and Lunar Laser Ranging experiments (dotted
  blue line). Vertical lines indicate the masses corresponding to the
  typical radii of the systems: 1AU (black solid line) and the orbital
  radii of the two binaries (dashed red and dot-dashed green
  lines). Right panel: upper bound on $\xi$ as a function of $m_s$.
  Linestyles are the same as in the left panel. Note that the
  theoretical bounds on the coupling parameters are $\omega>-3/2$ and
  $\xi<2$.}
\label{fig:bounds_all}
\end{figure*}

In addition to deriving the orbital period derivative due to
gravitational radiation, we also revisit the calculations of the
Shapiro time delay and of the Nordtvedt effect in the massive
Brans-Dicke theory. As we will see, the presence of the massive scalar
does not allow a straightforward implementation of the parametrized
post-Newtonian formalism. By comparing our results for the orbital
period derivative, Shapiro time delay and Nordtvedt parameter against
recent observational data, we put constraints on the parameters of the
theory: the scalar mass $m_s$ and the Brans-Dicke coupling parameter
$\omega_{\rm BD}$. Our bounds are summarized in Figure
\ref{fig:bounds_all}.

We find that the most stringent bounds come from the observations of
the Shapiro time delay in the Solar System provided by the Cassini
mission (these bounds were already studied by Perivolaropoulos,
although he used a slightly different notation
\cite{Perivolaropoulos:2009ak}). From the Cassini observations we
obtain $\omega_{\rm BD}>40000$ for $m_s<2.5\times10^{-20}\mathrm{eV}$,
to 95\% confidence. Observations of the Nordtvedt effect using the
Lunar Laser Ranging (LLR) experiment yield a slightly weaker bound of
$\omega_{\rm BD}>1000$ for $m_s<2.5\times10^{-20}\mathrm{eV}$.
Observations of the orbital period derivative of the circular
white-dwarf neutron-star (WD-NS) binary system PSR J1012+5307 yields
$\omega_{\rm BD}>1250$ for $m_s<10^{-20}\mathrm{eV}$. The limiting
factor here is our ability to obtain precise measurements of the
masses of the component stars as well as of the orbital period
derivative, once kinematic corrections have been accounted
for. However, there is considerably more promise in the eccentric
binary system PSR J1141-6545. This system has allowed for remarkably
precise measurements of the orbital period derivative, of the
component star masses and of the periastron shift, making it a
promising candidate for constraining alternative theories of
gravity. Unfortunately the system has nonnegligible
eccentricity. Generalizing our result for the orbital period
derivative to eccentric binaries is a significant (but worthy)
algebraic undertaking.

The plan of the paper is as follows. In section \ref{sec:mbd_theory}
we describe and motivate the Brans-Dicke theory with a massive scalar
field. In section \ref{sec:pn} we perform a post-Newtonian expansion
of the field equations. In section \ref{sec:shapiro_delay} we deal
with the Shapiro time delay. In section \ref{sec:eom} we proceed to
obtain the equations of motion of a binary system as well as the
periastron shift. In section \ref{sec:nordtvedt} we discuss the
Nordtvedt effect. In section \ref{sec:gw} we give details of the
derivation of the gravitational radiation damping of a compact binary
system due to scalar and tensor gravitational radiation. In section
\ref{sec:bounds} we use these results to put bounds on the parameters
of the theory. In the conclusions we point out possible future
extensions of our work. Appendix \ref{sec:pn_appendix} outlines a
step-by-step derivation of the post-Newtonian expansion of the scalar
field and of the metric.  Appendix \ref{sec:integration_appendix}
provides details on certain integrals that appear in the calculation
of the energy flux. Finally, Appendix \ref{sec:binary_appendix}
contains a short summary of compact binary observations relevant to
this work.

\section{The Brans-Dicke theory with a massive scalar field}\label{sec:mbd_theory}

\subsection{The generic scalar-tensor theory with a single scalar field}

A general class of scalar-tensor theories containing a single scalar
field in addition to the tensor field was studied by Bergmann and
Wagoner \cite{Bergmann:1968ve,Wagoner:1970vr}. We can characterize the
Bergmann-Wagoner theory via the following postulates:
%

\noindent
1) The principle of general covariance is imposed, leading to
  tensorial equations.

\noindent
2) The field equations are derived from the action
\begin{align}
S=\int\big(\mathscr{L}_\mathrm{G}+\mathscr{L}_\mathrm{M}\big)\;d^4x,
\end{align}
where $\mathscr{L}_\mathrm{G}$ and $\mathscr{L}_\mathrm{M}$ are the
Lagrangian densities for the gravitational and matter fields,
respectively.

\noindent
3) We postulate that the long-range forces of nature are mediated by
the three lowest spin bosons, and assume that the electromagnetic
field is the only vector field.  This leaves a scalar degree of
freedom $\phi$ and a tensor degree of freedom (the metric
$g_{\mu\nu}$) to describe the dynamics of the gravitational field.

\noindent
4) The field equations are of at most of second differential order,
and the tensor and scalar fields are nonminimally coupled; this leads
us to the general form
\begin{align}
\mathscr{L}_\mathrm{G}=(-g)^{\f{1}{2}}
\big[h(\phi)R+l(\phi)g^{\mu\nu}\phi_{,\mu}\phi_{,\nu}+\lambda(\phi)\big]
\end{align}
for the gravitational Lagrangian density, where $h(\phi)$, $l(\phi)$
and $\lambda(\phi)$ are arbitrary functions of the scalar field
$\phi$.

\noindent 
5) We postulate a principle of \emph{mutual coupling}, in which the
matter Lagrangian density depends on the gravitational fields
according to
\begin{align}
\mathscr{L}_\mathrm{M}=\mathscr{L}_\mathrm{M}(\psi^2(\phi)g^{\mu\nu},\Psi)\,,
\end{align}
where $\psi(\phi)$ is a fourth arbitrary function of $\phi$, and
$\Psi$ represents the collective matter fields. This guarantees
consistency with the strong equivalence principle \cite{Fujii:2003pa}.

Now let us make the conformal transformation
$g_{\mu\nu}\rightarrow\psi^2(\phi)g_{\mu\nu}$, and in doing so move
into a conformal frame in which the matter fields do not couple
directly (but only indirectly, via the metric) to the scalar field;
this is commonly referred to as the Jordan frame
\cite{Fujii:2003pa,Clifton:2011jh}.  Furthermore, without loss of
generality we can redefine the scalar field such that
$h(\phi)\rightarrow\phi$. These two redefinitions recast the action
into the form
\begin{eqnarray}
\label{action_general}
S&=&\f{1}{16\pi}\int
\left[
\phi R -\f{\omega(\phi)}{\phi} g^{\mu\nu}\phi_{,\mu}\phi_{,\nu}+M(\phi)
\right]
(-g)^\f{1}{2}d^4x \nonumber \\
&+&\int\mathscr{L}_\mathrm{M}(g^{\mu\nu},\Psi)d^4x,
\end{eqnarray}
which has the additional advantage that the resulting weak-field
equations for $g_{\mu\nu}$ and $\phi$ decouple from one another. The
generic theory now contains two undetermined functions: the
\emph{cosmological function} $M(\phi)$ and the \emph{coupling
  function} $\omega(\phi)$ (in the language of
\cite{Will:1993ns}). The effect of the coupling function on compact
binary dynamics has been studied extensively, and it can lead to
interesting consequences if ``spontaneous scalarization'' occurs
\cite{Damour:1992we,Damour:1993hw,Damour:1995kt,Damour:1996ke,Damour:1998jk}. Here
we focus on the cosmological function, which has three major effects
in the generic theory. Firstly, in the resulting field equations for
$g_{\mu\nu}$ it plays the role of a cosmological constant. Secondly,
it endows the scalar with mass: this manifests itself most clearly in
the fact that solutions for $\phi$ for isolated systems contain
Yukawa-like terms $e^{-m_sr}/r$, where $m_s$ is the mass of the scalar
field, which in turn gives the field a characteristic range $\ell
\sim1/m_s$ \cite{Will:1993ns}. Finally, the cosmological function may
introduce nonlinearities in the dynamics of the scalar field.

\subsection{The matter action and the field equations}
Let us now turn to the matter action. Throughout this paper we will
make the assumption that all bodies in our system can be treated as
point masses. Einstein, Infeld and Hoffmann (EIH)
\cite{Einstein:1938yz} developed a method for obtaining the equations
of motion for a system of gravitating point-like masses. In their
approach, one begins by obtaining the local gravitational field of a
single body (in a comoving frame), under the assumption that the body
is small and nearly spherical. One then proceeds to match the
interbody gravitational fields onto the obtained local field of the
single body under inspection; imposing self consistency yields the EIH
equations of motion. The same equations of motion can be obtained with
significantly less effort, albeit at the sacrifice of some rigor, by
taking the stress-energy tensor to be a distribution of
delta functions and neglecting any infinite self-energy terms as they
arise \cite{Will:1993ns}. In scalar-tensor theory, however, we must
deal with the additional complication that the inertial mass and
internal structure of a gravitating body will depend on the local
value of the scalar field (i.e. the local value of the effective
gravitational ``constant''). Variations in internal structure may act
back on the motion of the body, leading to violations of the (weak)
equivalence principle. Eardley \cite{1975ApJ...196L..59E} showed that
these effects could be accounted for by simply supposing that the
masses of the bodies are in general functions of the scalar field,
such that the matter action for a system of point-like masses can be
written as
\begin{align}
\label{matter_action}
S_\mathrm{M}=-\sum_a\int m_a(\phi)d\tau_a,
\end{align}
where the particles (labeled by $a$) have inertial masses $m_a(\phi)$,
and $\tau_a$ is the proper time of particle $a$ measured along its
worldline $x^\lambda_a$. The distributional stress-energy tensor
$T^{\mu\nu}$ and its trace $T$ hence take the form
\begin{align}
T^{\mu\nu}(x^\lambda)=(-g)^{-\f{1}{2}}\sum_am_a(\phi)
\frac{u^\mu u^\nu}{u^0}\delta^4(x^\lambda-x^\lambda_a)\,, \\
T=g_{\mu\nu}T^{\mu\nu}=-(-g)^{-\f{1}{2}}
\sum_a \frac{m_a(\phi)}{u^0}
\delta^4(x^\lambda-x_a^\lambda)\,.
\end{align}
Far from the system, the scalar will take on its cosmologically
imposed value, denoted by $\phi_0$. The relationship between the effective
gravitational constant, $G$, and the scalar field $\phi$ is therefore (in
our chosen system of units)
%
$G=\phi_0/\phi$.
%
In the post-Newtonian limit, we expand $\phi$ about its asymptotic
value and define the small perturbation $\varphi$ such that
$\phi=\phi_0+\varphi$. In this case, we can write the variation of the
inertial masses $m_a$ with $\phi$ as
\begin{align}
\label{mass_expansion}
m_a(\phi)= & m_a(\mathrm{ln}\;G)= m_a(\phi_0)
\left[1+s_a\Big(\f{\varphi}{\phi_0}\Big)\right. \nonumber \\
& -\left.
\f{1}{2}\big(s'_a-s_a^2+s_a)\Big(\f{\varphi}{\phi_0}\Big)^2+O\Big(\Big(\f{\varphi}{\phi_0}\Big)^3\Big)\right]\,,
\end{align}
where we have defined the ``first and second sensitivities'' $s_a$ and
$s'_a$ to be\footnote{White-dwarfs typically have sensitivities
  $s_a\sim10^{-4}$, neutron stars have sensitivities $s_a\sim0.2$, and
  black holes have $s_a=1/2$: see \cite{Zaglauer:1992bp} for detailed
  calculations.}
\begin{align}
\label{sensitivity}
s_a=-\f{\partial(\mathrm{ln}\;m_a)}{\partial(\mathrm{ln}\;G)}\Big|_{\phi_0}, \quad
s'_a=-\f{\partial^2(\mathrm{ln}\;m_a)}{\partial(\mathrm{ln}\;G)^2}\Big|_{\phi_0}\,.
\end{align}

The full action is now given by
\begin{align}
\label{action}
S= & \f{1}{16\pi}\int
\big[
\phi R -\f{\omega(\phi)}{\phi} g^{\mu\nu}\phi_{,\mu}\phi_{,\nu}+M(\phi)
\big](-g)^\f{1}{2}d^4x \nonumber \\
& -\sum_a\int m_a(\phi)d\tau_a.
\end{align}
By varying the action \eqref{action} with respect to the tensor and
scalar fields, respectively, we obtain the full field equations of the
generic theory described above:
\begin{widetext}
\begin{eqnarray}
\label{generic_tensor_fe}
R_{\mu\nu} - \frac{1}{2}g_{\mu\nu}R -\f{1}{2}\phi^{-1}M(\phi)g_{\mu\nu}&=& 8\pi\phi^{-1}T_{\mu\nu} 
 +\omega(\phi)\phi^{-2}(\phi_{,\mu}\phi_{,\nu}-\frac{1}{2}g_{\mu\nu}\phi_{,\alpha}\phi^{,\alpha})+\phi^{-1}(\phi_{,\mu\nu}-g_{\mu\nu}\Box_g\phi)\,,\\
%
%
\label{generic_scalar_fe}
\Box_g\phi+  \frac{\phi \frac{dM(\phi)}{d\phi}-2M(\phi)}{\big(3+2\omega(\phi)\big)} \nonumber 
&=&\f{1}{3+2\omega(\phi)}\Big(8\pi T^*-\f{d\omega(\phi)}{d\phi}\phi_{,\alpha}\phi^{,\alpha}\Big)\,,
\end{eqnarray}
\end{widetext}
where we have defined
%
$T^*=T-2\phi\f{\partial T}{\partial\phi}$
%
and $\Box_g$ is the curved space d'Alembertian, defined by
\begin{align}
\Box_g=(-g)^{-\frac{1}{2}}\partial_\nu((-g)^\frac{1}{2}g^{\mu\nu}\partial_\mu)\,.
\end{align}
A detailed derivation of this result can be found in
\cite{Fujii:2003pa}.

\subsection{Massive Brans-Dicke theory: The field equations and their weak-field limit}

As we recalled earlier, the effects of a generic coupling function on
the dynamics of compact binaries have been studied fairly extensively
by Damour and Esposito-Far\'ese
\cite{Damour:1992we,Damour:1993hw,Damour:1995kt,Damour:1996ke,Damour:1998jk}. Here
we are primarily interested in the effects of a nonzero mass of the
scalar field. In the limit where $m_s\to 0$, our final result for the
dipolar and quadrupolar flux can be shown to match Eq.~(6.40) in
\cite{Damour:1992we} (the monopole contribution vanishes for circular
orbits).

It would be interesting to study a theory with generic functional
forms for both $\omega(\phi)$ and $M(\phi)$, but for simplicity here
we will consider a constant coupling function:
$\omega(\phi)=\omega_{\rm BD}=\mathrm{constant}$, as in the usual
Brans-Dicke theory \cite{Brans:1961sx}. The scalar field equation then
reduces to
\begin{align}
\Box_g\phi+\f{\phi\f{dM(\phi)}{d\phi}-2M(\phi)}{3+2\omega_{\rm BD}}=\f{8\pi T^*}{3+2\omega_{\rm BD}}\,.
\end{align}
In order to get a handle on the effects of the cosmological function
$M(\phi)$, let us expand the metric about a Minkowski background
$\eta_{\mu\nu}$ and the scalar field around its (cosmologically
determined) constant background value $\phi_0$. Following closely the
method of \cite{Will:1993ns}, we define small perturbations $\varphi$,
$h_{\mu\nu}$ and $\theta_{\mu\nu}$ such that
\begin{align}
\label{perturb}
& \phi=\phi_0+\varphi, \nonumber \\
& g_{\mu\nu}=\eta_{\mu\nu}+h_{\mu\nu},\;g^{\mu\nu}=\eta^{\mu\nu}-h^{\mu\nu}, \nonumber \\
& \theta^{\mu\nu}=h^{\mu\nu}-\frac{1}{2}h\eta^{\mu\nu}-\Big(\frac{\varphi}{\phi_0}\Big)\eta^{\mu\nu}, 
\end{align}
Let us also expand $M(\phi)$ in a Taylor series about $\phi_0$:
\begin{align}
\label{Mphi}
M(\phi)=M(\phi_0)+M'(\phi_0)\varphi+\f{1}{2}M''(\phi_0)\varphi^2+\dots
\end{align}
We require that the expanded field equations are consistent at all
orders in $(v/c)^n$. Substituting the weak field perturbations
\eqref{perturb} into the field equations \eqref{generic_tensor_fe} and
\eqref{generic_scalar_fe} and examining the leading-order terms under
the assumption of asymptotic flatness, we find that
%
$M(\phi_0)=M'(\phi_0)=0$.
%
We are therefore left with the quadratic term, that endows the scalar
field with mass. To see this, let us substitute
$M(\phi)=\f{1}{2}M''(\phi_0)\varphi^2$ into the scalar field equation,
yielding
%
\begin{align}
\Box_g\phi-m_s^2(\phi-\phi_0)=\f{8\pi T^*}{3+2\omega_{\rm BD}}\,,
\end{align}
where we have defined the mass of the scalar field
\begin{align}
m_s^2\equiv 
-\f{\phi_0}{3+2\omega_{\rm BD}} M''(\phi_0)\,.
\end{align}
We will see shortly that $m_s$ is precisely the parameter appearing in
Yukawa-type corrections $\sim e^{-m_sr}$ to the Newtonian
gravitational potential, as well as the ordinary mass parameter in the
Klein-Gordon equation. Since the scalar field is expected to be small,
we will neglect cubic and higher-order terms in $M(\phi)$, that would
introduce additional nonlinearities into the scalar field equation.

In summary, with our choice of coupling and cosmological functions,
the field equations of the massive Brans-Dicke theory read
\begin{widetext}
\begin{eqnarray}
\label{mbd_tensor_fe}
&&R_{\mu\nu} - \frac{1}{2}g_{\mu\nu}R = - \f{3+2\omega_{\rm BD}}{4\phi_0\phi}m_s^2(\phi-\phi_0)^2g_{\mu\nu} + \frac{8\pi}{\phi}T_{\mu\nu}
 +\frac{\omega_{\rm BD}}{\phi^2}
\left(\phi_{,\mu}\phi_{,\nu}-\frac{1}{2}g_{\mu\nu}\phi_{,\alpha}\phi^{,\alpha}\right)+
\frac{1}{\phi}(\phi_{,\mu\nu}-g_{\mu\nu}\Box_g\phi)\,, \\
&&\Box_g\phi-m_s^2(\phi-\phi_0)=\f{8\pi T^*}{(3+2\omega_{\rm BD})}\,.
\label{mbd_scalar_fe}
\end{eqnarray}
\end{widetext}
\subsection{The weak-field limit}
Let us use the weak-field perturbations \eqref{perturb} to obtain the
field equations in the weak-field limit. Expanding the left hand side
of \eqref{mbd_tensor_fe} and imposing the harmonic gauge condition
$\theta^{\mu\nu}_{\;\;,\nu}=0$ we find
\begin{align}
R_{\mu\nu}-&\f{1}{2}g_{\mu\nu}R=-\f{1}{2}\Box_\eta\theta_{\mu\nu}+\f{\varphi_{,\mu\nu}}{\phi_0}-\eta_{\mu\nu}\Box_\eta\Big(\f{\varphi}{\phi_0}\Big)\,,
\end{align}
where $\Box_\eta$ is the flat-space d'Alembertian, and we neglected
quadratic and higher-order terms. The tensor field equation can hence
be written as
\begin{align}
\label{weak_tensor_fe}
\Box_\eta\theta^{\mu\nu}=-16\pi\tau^{\mu\nu}\,,
\end{align}
where
$\tau_{\mu\nu}=\phi_0^{-1}T_{\mu\nu}+t_{\mu\nu}$.
We have collected the quadratic and higher-order terms in the
perturbations $\varphi$ and $\theta_{\mu\nu}$ into the gravitational
stress-energy pseudotensor $t_{\mu\nu}$. By virtue of the gauge
condition on $\theta^{\mu\nu}$, we have the useful result that
\begin{align}
\tau^{\mu\nu}_{\;\;,\nu}=0\,.
\end{align}
Following a similar procedure for the scalar field equation, we expand
$\Box_g\phi$ in the weak-field perturbations:
\begin{align}
\Box_g\phi= & (1+\f{1}{2}\theta+\f{\varphi}{\phi_0})\Box_\eta\f{\varphi}{\phi_0}-\theta^{\mu\nu}\f{\varphi_{,\mu\nu}}{\phi_0}-\f{\varphi_{,\alpha}\varphi^{,\alpha}}{\phi_0^2} \nonumber \\
& +O(\theta^3,\;\theta^2\varphi,\varphi^2\theta\dots).
\end{align}
Substituting this back into the scalar field equation we find, as
anticipated, the standard Klein-Gordon equation
\begin{align}
\label{weak_scalar_fe}
(\Box_\eta-m_s^2)\varphi=-16\pi S\,,
\end{align}
where we have defined the source $S$ as
\begin{align}
\label{scalar_source}
S\equiv & -(6+4\omega_{\rm BD})^{-1}\Big(T-2\phi\frac{\partial T}{\partial\phi}\Big)\Big(1-\frac{1}{2}\theta-\frac{\varphi}{\phi_0}\Big) \nonumber \\
& -\frac{1}{16\pi}\big(\theta^{\mu\nu}\varphi_{,\mu\nu}+\phi_0^{-1}\phi_{,\alpha}\phi^{,\alpha}-m_s^2\phi_0^{-1}\varphi^2-\frac{1}{2}m_s^2\theta\varphi\big) \nonumber \\
& +O(\theta^3,\theta^2\varphi,\theta\varphi^2,\varphi^3)\,.
\end{align}

\section{Post-Newtonian expansion of the massive Brans-Dicke theory}\label{sec:pn} 

We will now perform a post-Newtonian expansion of the scalar and
tensor fields. This will allow us to
derive the Shapiro time delay (section \ref{sec:shapiro_delay}), the
equations of motion and periastron shift of compact binaries (section
\ref{sec:eom}) the Nordtvedt effect 
(section \ref{sec:nordtvedt}), and will be required for the derivation
the period derivative due to gravitational radiation (section \ref{sec:gw}). Before we proceed, it will be convenient to define
some auxiliary combinations containing $\omega_{\rm BD}$ that show up
repeatedly throughout the calculation:
\begin{align}
\label{parameter_definitions}
& \xi\equiv \f{1}{2+\omega_{\rm BD}}\,,\; \\
& \gamma\equiv \f{1+\omega_{\rm BD}}{2+\omega_{\rm BD}}\,,\; \\
& \alpha\equiv \f{1}{3+2\omega_{\rm BD}}\,.
\end{align}
Furthermore, for our choice of units the cosmologically imposed $\phi_0$
is given by
\begin{align}
\phi_0=\f{4+2\omega_\mathrm{BD}}{3+2\omega_\mathrm{BD}}.
\end{align}

Following very closely the method described in \cite{Will:1993ns} (see Appendix \ref{sec:pn_appendix} for details) we obtain
\begin{widetext}
\begin{eqnarray}
\label{pn_phi}
\frac{\phi}{\phi_0} &=& \xi\sum_a \frac{m_a}{r_a}(1-2s_a)e^{-m_sr_a}
 +\xi^2\sum_{a\neq b}\frac{m_a m_b}{r_a r_{ab}}(s_a+2s_a'-2s_a^2)
 \times e^{-m_sr_a}(1-2s_b)e^{-m_sr_{ab}} \nonumber \\
 &+&\frac{1}{2}\xi^2\sum_{a, b}\frac{m_a m_b}{r_a r_{b}}
 \times (1-2s_a)e^{-m_sr_a}(1-2s_b)e^{-m_sr_b}
 -\xi\sum_{a\neq b} \frac{m_a m_b}{r_a r_{ab}} (1-2s_a) e^{-m_sr_a}\phi_0^{-1}
 \times \Big(1+\alpha(1-2s_b)e^{-m_sr_{ab}}\Big) \nonumber \\
 &-&\frac{1}{2}\xi\sum_a \frac{m_a v_a^2}{r_a}(1-2s_a)e^{-m_sr_a}
 -\frac{1}{2}\xi\sum_a\frac{\partial^2}{\partial t^2}
 \Big(\frac{e^{-m_sr_a}}{m_s}\Big)(1-2s_a) +O(6)\,, \\
g_{00} &=& -1+2\phi_0^{-1}\sum_a\frac{m_a}{r_a}\Big(1+\alpha(1-2s_a)e^{-m_sr_a}\Big)
-2\phi_0^{-2}\sum_{a, b}\frac{m_a m_b}{r_a r_b}\Big(1+\alpha(1-2s_a) e^{-m_sr_a}\Big) 
\times\Big(1+\alpha(1-2s_b)e^{-m_sr_b}\Big) \nonumber \\
&-&2\sum_{a\neq b}\frac{m_a m_b}{r_a r_{ab}}
\Big[\phi_0^{-2}\Big(1+\alpha e^{-m_sr_a}\Big) 
\times\Big(1+\alpha(1-2s_b)e^{-m_sr_{ab}}\Big) 
-\xi\phi_0^{-1}s_a(e^{-m_sr_a}+(1-2s_b)e^{-m_sr_{ab}}) \nonumber \\
&-&\xi^2(s_a+2s_a'-2s_a^2)(1-2s_b)e^{-m_sr_a}e^{-m_sr_{ab}}\Big]
-\sum_a\frac{m_av_a^2}{r_a}\Big[1+2\gamma+\xi s_ae^{-m_sr_a} 
+\frac{1}{2}\xi(1-e^{-m_sr_a})\Big] \nonumber \\
&+&\sum_am_a\xi(1-2s_a) 
\times\frac{\partial^2}{\partial t^2}\Big(\frac{(2+m_sr_a)(1-e^{-m_sr_a})-2m_sr_a}{2m_s^2r_a}\Big)  + O(6)\,,
\label{pn_g00} \\
g_{0i} &=& -2(1+\gamma)\sum_a\frac{m_av_a^i}{r_a}-\frac{1}{2}\sum_am_a\phi_0^{-1}
\times\frac{\partial^2}{\partial t\partial x^i}\Big(r_a+2\alpha(1-2s_a)\frac{e^{-m_sr_a}+m_sr_a-1}{m_s^2r_a}\Big) +O(5)\,, \label{pn_g0i} \\
g_{ij} &=& \delta_{ij}+2\phi_0^{-1}\delta_{ij}\sum_a\frac{m_a}{r_a}\Big(1-\alpha(1-2s_a)e^{-m_sr_a}\Big) + O(4)\,. \label{pn_gij}
\end{eqnarray}
\end{widetext}
In the limit $m_s\rightarrow0$, the above results reduce to those
obtained in the massless Brans-Dicke case \cite{Will:1989sk}.

Substituting these results into \eqref{scalar_source} we find an
expression for the source $S$ in the near zone, to $O(2)$:
\begin{widetext}
\begin{eqnarray}
S(x^\lambda)=\f{\alpha}{2}
\sum_a m_a\Big[(1-2s_a)\Big(1-\f{1}{2}v_a^2-\frac{1}{\phi_0}
\sum_{b}\f{m_b}{r_{b}}\Big)
 + \xi(2s_a'-2s_a^2+2s_a-\f{1}{2})\sum_{b}\f{m_b (1-2s_b)}{r_{b}}e^{-m_s r_{b}}
\Big]
\delta^4(x^\lambda-x_a^\lambda)
\nonumber
\end{eqnarray}
\end{widetext}

\section{Shapiro time delay}\label{sec:shapiro_delay}

Using the post-Newtonian expansion of the metric, we can derive an
expression for the Shapiro time delay of a light ray passing near a
massive body. We note first that the parametrized post-Newtonian (PPN)
formalism is not viable when dealing with theories that contain
massive fields. In fact, Newtonian order terms are modified by the
presence of massive fields, in the sense that the Newtonian potential
acquires a Yukawa-like correction of the form
\begin{align}
\tilde{U}(\mathbf{x},t)=\f{1}{\phi_0}\int\f{\rho(\mathbf{x}',t)}{|\mathbf{x}-\mathbf{x}'|}(1+\alpha e^{-m_s|\mathbf{x}-\mathbf{x}'|})d^3\mathbf{x}'.
\end{align}
The impact of this fact for our current purpose is significant: the
above potential cannot be expanded in powers of $1/r$, and the
coefficients of modified post-Newtonian potentials in the
post-Newtonian metric are not constants, but they have a spatial
dependence. Nonetheless, we can use the derived metric to obtain an
expression for the equations of motion of a photon, and use this to
obtain an expression for the Shapiro delay. We will follow closely the
method described in \cite{Will:1993ns}. A similar calculation was
carried out by Perivolaropoulos \cite{Perivolaropoulos:2009ak}; he
used a different definition of the mass of the scalar field, but his
results are consistent with those derived here. 

For a photon traveling along a null geodesic,
\begin{align}
g_{\mu\nu}u^\mu u^\nu=0. 
\end{align}
To requisite order, $O(2)$, the equation of motion can be written as
\begin{align}
\label{photon_eom_O2}
-1+h^{(2)}_{00}+(\delta_{ij}+h_{ij}^{(2)})u^iu^j=0\,,
\end{align}
where $h_{\mu\nu}^{(n)}$ is the $O(n)$ order correction to the
metric. Specializing to a single spherically symmetric source of mass
$M$ (and negligible sensitivity) at the origin, the post-Newtonian
corrections to the metric are (from equations \eqref{pn_g00} and
\eqref{pn_gij})
\begin{align}
\label{O2_corrections}
& h_{00}^{(2)}=2\phi_0^{-1}\f{M}{r}\big(1+\alpha e^{-m_sr}\big)=2\tilde{U}\,, \nn \\
& h_{ij}^{(2)}=2\phi_0^{-1}\f{M}{r}\big(1-\alpha e^{-m_sr}\big)\delta_{ij}=2U(1-\alpha e^{-m_sr})\delta_{ij}\,.
\end{align}
%

Substituting these into \eqref{photon_eom_O2}, the equation of motion
for the photon now reads
\begin{align}
-1+2\tilde{U}+\left(1+2(1-\alpha e^{-m_sr})U\right)|\mathbf{u}|^2=0\,.
\end{align}
The unperturbed Newtonian trajectory of the photon will simply be
$x^i(t)=x_e^i+n^i(t-t_e)$, where the photon is emitted from
$\mathbf{x_e}$ in direction $\mathbf{n}$ at time $t_e$. Let us now
parametrize the post-Newtonian correction to the trajectory by
$x_{\rm PN}^i(t)$, where the corrected trajectory is then given by
$x^i(t)=x_e^i+n^i(t-t_e)+x_{\rm PN}^i(t)$. Substituting this into the
above, we find that the post-Newtonian correction to the trajectory
satisfies
\begin{align}
\mathbf{n}\cdot\f{d\mathbf{x}_{\rm PN}}{dt}=\f{dx_{\rm PN}^{\parallel}}{dt}=-2U\,.
\end{align}
Integrating with respect to time, we obtain
\begin{align}
x^\parallel_{\rm PN}(t)=-2\int_{t_e}^tU dt'\,.
\end{align}
The time taken for the photon to travel from $\mathbf{x}_e$ to some
other point $\mathbf{x}$ and back again is hence given by
\begin{align}
\label{shapiro_delay}
\Delta t=2|\mathbf{x}-\mathbf{x}_e|+4\int_{t_e}^tU dt'\,.
\end{align}
The travel time correction $\delta t$ due to the Shapiro delay
corresponds to the second term on the right-hand side. Performing the
integration, we find for the Shapiro delay term
\begin{align}
\label{shapiro_delay_limits}
\delta t=4M\;\mathrm{ln}
\Big[\f{(r_e+\mathbf{r}_e\cdot\mathbf{n})
(r_p-\mathbf{r}_p\cdot\mathbf{n})}{r_b^2}\Big]\,,
\end{align}
where the photon is emitted from $\mathbf{r}_e$ in direction
$\mathbf{n}$, travels to $\mathbf{r}_p$ and back again, $M$ is the
mass of the body causing the time-delay and $r_b$ is the impact
parameter of the photon with respect to the source. The mass appearing
in \eqref{shapiro_delay_limits} is not a measurable quantity; what is
actually measured is the Keplerian mass $M_\mathrm{K}=M(1+\alpha
e^{-m_sr})$, where $r$ should be thought of as a fixed quantity which
depends on how the Keplerian mass of the body was determined.
In terms of $M_\mathrm{K}$ we have
\begin{align}
\label{shapiro_delay_gammatilde}
\delta t&= \f{4M_\mathrm{K}}{1+\alpha e^{-m_sr}}\;\mathrm{ln}\Big[\f{(r_e+\mathbf{r}_e\cdot\mathbf{n})(r_p-\mathbf{r}_p\cdot\mathbf{n})}{r_b^2}\Big] \nn \\
& =2(1+\tilde{\gamma})M_\mathrm{K}\;\mathrm{ln}\Big[\f{(r_e+\mathbf{r}_e\cdot\mathbf{n})(r_p-\mathbf{r}_p\cdot\mathbf{n})}{r_b^2}\Big],
\end{align}
where in the second line we have defined
\begin{align}
\label{gammatilde}
\tilde{\gamma} = \f{1-\alpha e^{-m_sr}}{1+\alpha e^{-m_sr}}\,.
\end{align}
In the case of the solar system, the $r$ appearing in the definition
of $\tilde{\gamma}$ should be set to $1 \mathrm{AU}$, since this is
the scale associated with the determination of the Keplerian mass of
the Sun.
In any metric theory of gravity where the PPN formalism can be applied
in a straightforward manner, the obtained expression for the Shapiro
delay is identical to \eqref{shapiro_delay_gammatilde}, only with
$\tilde{\gamma}$ replaced by the PPN parameter $\gamma$ (see for
example \cite{Will:1993ns}).  We can therefore compare
$\tilde{\gamma}$ directly with the observational constraints on
$\gamma$ from Shapiro time delay measurements to obtain an exclusion
region in the $(\omega_{\rm BD},m_s)$-plane. In section
\ref{sec:shapiro_bounds} we will do precisely this, comparing the
derived expression for $\tilde{\gamma}$ to the constraints on $\gamma$
from time-delay measurements obtained by the Cassini mission.

Note that in the limit where $m_s\rightarrow\infty$,
$\tilde{\gamma}\rightarrow 1$, i.e. the GR value of the PPN parameter
$\gamma$. In the limit where $m_s\rightarrow 0$ we have instead
$\tilde{\gamma}\rightarrow \gamma = (1+\omega_{\rm BD})/(2+\omega_{\rm
  BD})$, i.e. the value of $\gamma$ in the massless Brans-Dicke
theory.

\section{Equations of motion and periastron advance} \label{sec:eom}

Armed with the post-Newtonian expansion of the fields, we are now in a
position to obtain the EIH equations of motion. From
\eqref{matter_action}, the matter Lagrangian for the $a$th body in the system is
given by
\begin{align}
\label{lagrangian}
L_a = m_a(\phi)\left(-g_{00}-2g_{0i}v_a^i-g_{ij}v_a^iv_a^j\right)^\f{1}{2}\,.
\end{align}
To obtain an $n$-body action we follow the procedure detailed after
Eq.~(11.90) of \cite{Will:1993ns}. We substitute the post-Newtonian
expressions for the metric and scalar fields obtained in the previous
section and use the expansion of $m_a(\phi)$ in
\eqref{mass_expansion}. We first make the gravitational terms in
$L_a$ manifestly symmetric under interchange of all pairs
of particles, then we take one of each such term generated in
$L_a$, and sum over $a$. To $O(4)$ we find
\begin{align}
\label{nbody_lagrangian}
L_\mathrm{EIH}= & -\sum_am_a\big(1-\f{1}{2}v_a^2-\f{1}{8}v_a^4\big) \nonumber \\
& + \f{1}{2}\sum_{a\neq b}\f{m_am_b}{r_{ab}}\Big[\mathcal{G}_{ab}+3\mathcal{B}_{ab}v_a^2-\sum_{c\neq a}\mathcal{D}_{abc}\f{m_c}{r_{ac}} \nonumber \\
& -\f{1}{2}(\mathcal{G}_{ab}+6\mathcal{B}_{ab})\mathbf{v}_a\cdot \mathbf{v}_b-\f{1}{2}\mathcal{G}_{ab}(\mathbf{v}_a\cdot \mathbf{n}_{ab})(\mathbf{v}_b\cdot \mathbf{n}_{ab})\Big]\,,
\end{align}
where we have defined
\begin{align}
& \mathbf{n}_{ab}=\f{\mathbf{r}_a-\mathbf{r}_b}{r_{ab}}\,, \nonumber \\
& \mathcal{G}_{ab}=1-\f{1}{2}\xi\big[1-(1-2s_a)(1-2s_b)e^{-m_sr_{ab}}\big]\,, \nonumber \\
& \mathcal{B}_{ab}=\f{1}{3}(2\gamma+1)+\f{1}{6}\xi\big[1-(1-2s_a)(1-2s_b)e^{-m_sr_{ab}}\big]\,,
\end{align}
and
\begin{align}
\mathcal{D}_{abc}= & 1-\f{1}{2}\xi[2-(1-2s_a)(1-2s_b)e^{-m_sr_{ab}} \nonumber \\
& -(1-2s_a)(1-2s_c)e^{-m_sr_{ac}}] \nonumber \\
& +\f{1}{4}\xi^2\Big[1-(1-2s_a)(1-2s_b)e^{-m_sr_{ab}} \nonumber \\
& -(1-2s_a)(1-2s_c)e^{-m_sr_{ac}} \nonumber \\
& +\big(1-4(s_a+s_a'-s_a^2)\big) \nonumber \\
& \times(1-2s_b)(1-2s_c)e^{-m_sr_{ab}}e^{-m_sr_{ac}}\Big].
\end{align}
Now let us now specialize to a two-body system with the center of mass
at the origin; to this end let us define
\begin{align}
\label{two-body}
& \mathbf{r}=\mathbf{r}_2-\mathbf{r}_1\,,\;m=m_1+m_2\,, \nonumber \\
& \;\delta m=m_2-m_1\,,\;\mu=\f{m_1m_2}{m}\,.
\end{align}
We also write $\mathcal{G}_{12}=\mathcal{G}$ and
$\mathcal{B}_{12}=\mathcal{B}$. With this specialization made, the
equations of motion are found to be
\begin{align}
\label{eom}
\mathbf{a}= & -\f{m\mathbf{r}}{r^3}\Big[\tilde{\mathcal{G}}-3\tilde{\mathcal{G}}\mathcal{B}\f{m}{r}-\f{1}{2}\big(\tilde{\mathcal{G}}-3\tilde{\mathcal{B}}\big)v^2 \nonumber \\
& -\f{1}{2}\big(\mathcal{D}_{211}+\tilde{\mathcal{D}}_{211}\big)\f{m_1}{r}-\f{1}{2}\big(\mathcal{D}_{122}+\tilde{\mathcal{D}}_{122}\big)\f{m_2}{r} \nonumber \\
& -2\mathcal{G}\tilde{\mathcal{G}}\f{\mu}{r}+\big(\mathcal{G}+2\tilde{\mathcal{G}}\big)\f{\mu}{m}v^2-\f{1}{2}\big(4\mathcal{G}-\tilde{\mathcal{G}}\big)\f{\mu}{m}(\mathbf{v\cdot n})^2\Big] \nonumber \\
& +\f{m(\mathbf{r\cdot v})\mathbf{v}}{r^3}\Big[\tilde{\mathcal{G}}+3\mathcal{B}+(\mathcal{G}-3\tilde{\mathcal{G}})\f{\mu}{m}\Big]\,,
\end{align}
where
\begin{align}
\label{G_and_B_definitions}
& \tilde{\mathcal{G}}=1-\f{1}{2}\xi[1-(1-2s_1)(1-2s_2)(1+m_sr)e^{-m_sr}]\,, \nonumber \\
& \tilde{\mathcal{B}}=\f{1}{3}(2\gamma+1)+\f{1}{6}\xi[1-(1-2s_1)(1-2s_2)(1+m_sr)e^{-m_sr}]
\end{align}
and
\begin{align}
\tilde{\mathcal{D}}_{122}= & 1-\xi[1-(1-2s_1)(1-2s_2)(1+m_sr)e^{-m_sr}] \nonumber \\
& +\f{1}{4}\xi^2\Big[1-2(1-2s_1)(1-2s_2)(1+m_sr)e^{-m_sr} \nonumber \\
& +(1-4(s_1+s_1'-s_1^2))(1-2s_2)^2(1+2m_sr)e^{-2m_sr}\Big].
\end{align}

\subsection{Periastron advance\label{sec:periastron}}

With the equation of motion in hand, we can view the post-Newtonian
corrections together with the scalar Yukawa-like terms as
perturbations of the Keplerian orbit and employ the method of
osculating elements \cite{Will:1993ns} to obtain an expression for the
periastron advance of the binary system. In contrast to the massless
Brans-Dicke case (treated in \cite{Will:1993ns}), the integrals that
appear in this perturbation expansion cannot be written in closed
form, so an expansion in powers of the eccentricity $e$ is required to
obtain closed-form expressions. Fortunately for our current purposes
we will only require the result in the two limiting cases of very
light and very massive scalars. In the former limit $m_sr\ll 1$, the
periastron advance reduces to the massless Brans-Dicke result
\cite{Will:1993ns}
\begin{align}
\label{periastron_advance}
\dot{\omega}=\f{6\pi m}{a(1-e^2)P}\mathcal{P}\mathcal{G}^{-1}\,,
\end{align}
where $a$ and $e$ are the semi-major axis and eccentricity, $P$ is the
period and $\mathcal{P}$ is given by
\begin{align}
\mathcal{P}=\mathcal{GB}+\f{1}{6}\mathcal{G}^2-\f{1}{6}\f{m_1\mathcal{D}_{211}+m_2\mathcal{D}_{122}}{m}\,. 
\end{align}
In the limit of a very massive scalar $m_sr\gg 1$ the expression for
the periastron advance reduces instead to the familiar GR result:
\begin{align}
\dot{\omega}=\f{6\pi m}{a(1-e^2)P}\mathcal{G}\,.
\end{align}

\section{Nordtvedt effect}\label{sec:nordtvedt}
%

Scalar-tensor theories of gravity predict that massive bodies with a
significant amount of gravitational self-energy do not follow
geodesics of the background metric; in fact, massive bodies with
different gravitational self-energies will follow different
trajectories, leading to direct violation of the strong equivalence
principle. This is known as the Nordtvedt effect, and leads to
detectable effects in the Solar System. Most notably, it leads to a
polarization of the Moon's orbit around the Earth
\cite{0034-4885-45-6-002, Will:1993ns}, which can be constrained using
lunar ranging experiments. Let us look at how this effect arises in
the massive Brans-Dicke theory.

The effect is usually parametrized by the Nordtvedt parameter
$\eta_\mathrm{N}$, which can be determined directly from the PPN
metric of a given theory, and it turns out to be some simple
combination of PPN parameters. However, as we have seen previously, in
the case of the massive Brans-Dicke theory the PPN formalism is not
directly applicable. However we can extract an ``effective'' Nordtvedt
parameter from the equations of motion. To do this, let us consider
the relative acceleration of a pair of bodies $A$ and $B$,
$\mathbf{a}_\mathrm{AB}=\mathbf{a}_\mathrm{A}-\mathbf{a}_\mathrm{B}$,
in the field of a third body $C$, with $r_\mathrm{AB}\ll
r_\mathrm{AC}$ and $r_\mathrm{AC}\simeq r_\mathrm{BC}$. The Nordtvedt
effect will result in an anomalous difference in the accelerations of
$A$ and $B$ towards $C$, proportional to the difference in the specific
gravitational self-energies of the two bodies $A$ and $B$
\cite{Nordtvedt:1968qs, 0034-4885-45-6-002, Will:1993ns}. Since the
sensitivity $s_a$ of a body is related to its gravitational
self-energy $\Omega_a$ by $s_a=\Omega_a/m_a$ (in the weak field
limit), the extra term arising in $\mathbf{a}_\mathrm{AB}$ due to the
Nordtvedt effect will be proportional to the the difference in
sensitivities $\mathcal{S}=s_\mathrm{B}-s_\mathrm{A}$.

To Newtonian order, the $n$-body Lagrangian \eqref{nbody_lagrangian} is
given by
\begin{align}
L_\mathrm{EIH}=-\sum_a m_a(1-\f{1}{2}v_a^2)+\f{1}{2}\sum_{a\neq b}\f{m_a m_b}{r_{ab}}\mathcal{G}_{ab},
\end{align}
and the $n$-body equations of motion are hence
\begin{align}
\mathbf{a}_a&=-\sum_{b\neq a}\f{m_b}{r_{ab}^2}\mathcal{G}_{ab}\hat{\mathbf{r}}_{ab} \\
&-\f{1}{2}\sum_{b\neq
  a}\f{m_b}{r_{ab}^2}\xi(1-2s_a)(1-2s_b)m_sr_{ab}e^{-m_sr_{ab}}\hat{\mathbf{r}}_{ab}\,. \nn
\end{align}
The relative acceleration of two bodies $A$ and $B$ in the field of a
third body $C$ is then

\begin{align}
\mathbf{a}_\mathrm{AB}&=\mathbf{a}_\mathrm{B}-\mathbf{a}_\mathrm{A} \\
&=\f{\mathcal{G}_\mathrm{AB}(m_\mathrm{A}+m_\mathrm{B})}
{r_\mathrm{AB}^2}\hat{\mathbf{r}}_\mathrm{AB}-\f{\mathcal{G}_\mathrm{BC}m_\mathrm{C}}
{r_\mathrm{BC}^2}\hat{\mathbf{r}}_\mathrm{BC}+\f{\mathcal{G}_\mathrm{AC}m_\mathrm{C}}
{r_\mathrm{AC}^2}\hat{\mathbf{r}}_{AC} \nn \\
&+\f{1}{2}\left(\f{m_\mathrm{A}+m_\mathrm{B}}
{r_\mathrm{AB}}\xi(1-2s_\mathrm{A})(1-2s_\mathrm{B})m_se^{-m_sr_\mathrm{AB}}\right)\hat{\mathbf{r}}_\mathrm{AB} \nn \\
&-\f{1}{2}\f{m_\mathrm{C}}{r_\mathrm{BC}^2}
\xi(1-2s_\mathrm{B})(1-2s_\mathrm{C})m_sr_\mathrm{BC}e^{-m_sr_\mathrm{BC}}\hat{\mathbf{r}}_\mathrm{BC} \nn \\
&+\f{1}{2}\f{m_\mathrm{C}}{r_\mathrm{AC}^2}
\xi(1-2s_\mathrm{A})(1-2s_\mathrm{C})m_sr_\mathrm{AC}e^{-m_sr_\mathrm{AC}}\hat{\mathbf{r}}_\mathrm{AC}\,. \nn
\end{align}
Regrouping terms together appropriately and assuming that
$r_\mathrm{AB}\ll r_\mathrm{AC}$, $r_\mathrm{AC}\simeq r_\mathrm{BC}$,
we can rewrite this as
\begin{align}
\mathbf{a}_\mathrm{AB}&=
-\f{m^*\hat{\mathbf{r}}_\mathrm{AB}}
{r_\mathrm{AB}^2}+\f{1}{\phi_0}\left(\f{m_\mathrm{C}\hat{\mathbf{r}}_\mathrm{AC}}
{r_\mathrm{AC}^2}-\f{m_\mathrm{C}\hat{\mathbf{r}}_\mathrm{BC}}{r_\mathrm{BC}^2}\right) \\
&+\left[\xi(1-2s_\mathrm{C})(1+m_sr_\mathrm{AC})e^{-m_sr_\mathrm{AC}}\right](s_\mathrm{B}-s_\mathrm{A})
\f{m_\mathrm{C}\hat{\mathbf{r}}_\mathrm{AC}}{r_\mathrm{AC}^2} \nn \\
&=-\f{m^*\hat{\mathbf{r}}_\mathrm{AB}}{r_\mathrm{AB}^2}+\f{1}{\phi_0}
\left(\f{m_\mathrm{C}\hat{\mathbf{r}}_\mathrm{AC}}{r_\mathrm{AC}^2}-\f{m_\mathrm{C}\hat{\mathbf{r}}_\mathrm{BC}}
{r_\mathrm{BC}^2}\right)+\eta_\mathrm{N}\mathcal{S}\f{m_\mathrm{C}\hat{\mathbf{r}}_\mathrm{AC}}{r_\mathrm{AC}^2}\,,
\nn
\end{align}
where the first term is the Newtonian acceleration between the two
bodies, the second term is the tidal correction to the orbit of the
system $(A,B)$ and the final term (proportional to
$\mathcal{S}=s_\mathrm{B}-s_\mathrm{A}$) is the difference in the
accelerations of $A$ and $B$ towards the third body $C$ due to the
Nordtvedt effect (cf.~\cite{Will:1993ns}). In the second line we have
rewritten the third term in the conventional form from which the
Nordtvedt parameter is usually defined; we can then simply read off
the effective Nordtvedt parameter
\begin{align}
\label{nordtvedt_effective}
\eta_\mathrm{N}=\xi(1+m_sr)(1-2s_\mathrm{C})e^{-m_sr}\,,
\end{align}
where $r$ is now taken to be the distance from $C$ to the system
$(A,B)$. Note that if the Sun were replaced by a black hole ($s_C =
1/2$), there would be no Earth-Moon Nordtvedt effect.  In section
\ref{sec:nordtvedt_bounds} we will compare the effective
$\eta_\mathrm{N}$ to the measured value of the Nordtvedt parameter
provided by Lunar Laser Ranging experiments to obtain bounds on
$(\omega_{\rm BD},m_s)$.

\section{Gravitational radiation from compact binaries}\label{sec:gw}
\subsection{Tensor radiation}

In this section we will follow very closely the general method
described in \cite{Will:1993ns}. The power radiated in gravitational
waves due to tensor radiation in the Brans-Dicke theory is given by
%
\begin{align}
\dot{E}=-\f{R^2}{32\pi}\phi_0\Big\langle\oint\theta^{ij}_\mathrm{TT,0}\theta^{ij}_\mathrm{TT,0}d\Omega\Big\rangle\,,
\end{align}
where the angular brackets represent an average over one orbital
period and $\theta^{ij}_\mathrm{TT}$ is the transverse-traceless (TT) part
of $\theta^{ij}$.

In order to obtain a formal solution to the linearized tensor wave
equation \eqref{weak_tensor_fe}, we simply fold the source
$\tau_{\mu\nu}$ with the retarded Green's function of the flat-space
d'Alembertian operator
\begin{align}
G(t-t',\mathbf{R-r'})=\f{\delta(t-t'-|\mathbf{R-r'}|)}{|\mathbf{R-r'}|}\,,
\end{align}
with the result
\begin{align}
\label{weak_tensor_general_sol}
\theta^{\mu\nu}(t,\mathbf{R})=4\int_\mathscr{N}\f{\tau^{\mu\nu}(t-|\mathbf{R-r'}|,\mathbf{r'})}{|\mathbf{R-r'}|}d^3\mathbf{r'}\,.
\end{align}
Here the integral over $t'$ has been carried out immediately, and the
spatial integration region $\mathscr{N}$ is over the near zone.  If we
make the assumption that the field point in is the radiation zone,
such that $|\mathbf{r}'|\ll|\mathbf{R}|$, and make the slow-motion
approximation, we can expand the $\mathbf{r'}$ dependence of the
integrand and write
\begin{align}
\theta^{\mu\nu}=\f{4}{R}\sum_{m=0}^\infty\f{1}{m!}\f{\partial^m}{\partial t^m}\int_\mathscr{M}\tau^{\mu\nu}(t-R,\mathbf{r'})(\mathbf{n\cdot r'})^md^3\mathbf{r'}\,,
\end{align}
where $\mathbf{n}=\mathbf{R}/R$, and the integration is now over $\mathcal{M}$, which is the intersection 
of the world tube of the near zone with the constant time hypersurface $t_\mathcal{M}=t-R$
\cite{Will:1996zj}. For the purpose of obtaining the
power loss due to gravitational radiation, we are ultimately
interested in $\theta^{\mu\nu}_{\;\;,0}$. Due to our choice of gauge
$\theta^{00}_{\;\;,0}=\theta^{\mu0}_{\;\;,0}=0$, and hence we only
require the spatial components $\theta^{ij}$, which are given (to
leading order) by
\begin{align}
\label{tau_multipole}
\theta^{ij} & = \f{4}{R}\int\tau^{ij}(t-R,\mathbf{r'})d^3\mathbf{r'} \nonumber \\
& =\f{2}{R}\f{\partial^2}{\partial t^2}\int\tau^{00}(t-R,\mathbf{r'})r'^{i}r'^{j}d^3\mathbf{r'}\,.
\end{align}
Here we have written the monopole moment of $\tau^{ij}$ as the time
derivative of the quadrupole moment of $\tau^{00}$, by exploiting the
conservation law $\tau^{\mu\nu}_{\;\;,\nu}$ together with the
slow-motion approximation. There can be no contribution from the
dipole moment of $\tau^{00}$ in \eqref{tau_multipole} to order
$O(\f{mv}{R})$, since the time derivative $\f{\partial x}{\partial
  t}\sim v$. The quadrupole moment of $\tau^{ij}$ only comes in at
higher order, and hence we only require the leading-order contribution
from $\tau^{00}$:
\begin{align}
\tau^{00}=\f{1}{2}(1+\gamma)\sum_am_a\delta^3(\mathbf{r'-r_a})\,.
\end{align}
Substituting this into \eqref{tau_multipole} we obtain
\begin{align}
\theta^{ij}=(1+\gamma)R^{-1}\f{d^2}{d t^2}\sum_a m_a r_a^i r_a^j\,.
\end{align}
Specializing to a two-body system with the center of mass at the
origin using \eqref{two-body}, we obtain to the requisite order
\begin{align}
\label{theta_harmonic}
\theta^{ij}(t,\mathbf{R})=2(1+\gamma)R^{-1}\mu\Big(v^iv^j-\tilde{\mathcal{G}} m\f{r^i r^j}{r^3}\Big)\,,
\end{align}
where we have used \eqref{eom} to replace $\ddot{r}^i$ (to leading
order) where necessary.  

We now need to project \eqref{theta_harmonic} onto the TT gauge by
applying the projector 
\begin{eqnarray}
\Lambda(\mathbf{\hat{n}})_{ij,kl}&=& \delta_{ik}\delta_{jl}-\f{1}{2}\delta_{ij}\delta_{kl}-n_in_k\delta_{jl} \nonumber \\
&+&\f{1}{2}n_kn_l\delta_{ij}+\f{1}{2}n_in_j\delta_{kl}+\f{1}{2}n_in_jn_kn_l\,,
\end{eqnarray}
which satisfies
%
$\Lambda_{ij,kl}\Lambda_{kl,nm}=\Lambda_{ij,nm}$
%
\cite{Maggiore:1900zz} to $\theta^{kl}$:
\begin{align}
\theta^{ij}_\mathrm{TT}=\Lambda(\mathbf{\hat{n}})_{ij,kl}\theta^{kl}\,.
\end{align}
The result is
\begin{align}
\label{tensor_power_loss_harmonic}
\dot{E}=
-\f{R^2}{32\pi}\phi_0\Big\langle
\oint\Lambda_{ij,kl}\theta^{ij}_{,0}\theta^{ij}_{,0}d\Omega\Big\rangle\,.
\end{align}
We now note that the only $\mathbf{\hat{n}}$ dependence in the
integrand of \eqref{tensor_power_loss_harmonic} in contained in the
$\Lambda_{ij,kl}$. Performing the integral over the solid angle we find
\begin{align}
\oint\Lambda_{ij,kl}d\Omega=\f{2\pi}{15}\big(11\delta_{ik}\delta_{jl}-4\delta_{ij}\delta_{kl}+\delta_{il}\delta_{jk}\big)\,,
\end{align}
where we have used the identity
\begin{align}
\label{unit_int_identity}
\oint n^{i_1}n^{i_2}\cdots n^{i_{2l}}d\Omega=\frac{4\pi\delta^{(i_1i_2}\delta^{i_3i_4}\cdots\delta^{i_{2l-1}i_{2l})}}{(2l+1)!!}\,.
\end{align}
Substituting this result back into \eqref{tensor_power_loss_harmonic}
we obtain
\begin{align}
\dot{E}=-\f{R^2}{32\pi}\phi_0\f{2\pi}{15}\Big\langle12\;\theta^{ij}_{\;\;,0}\theta^{ij}_{\;\;,0}-4\;\theta^i_{\;i,0}\theta^i_{\;i,0}\Big\rangle\,.
\end{align}
At this point we will specialize to a circular orbit, which we will
parametrize by
\begin{align}
\label{circle}
& r_1=r\;\mathrm{cos}\big(\omega (t-R)\big)\,,\; \\
& r_2=r\;\mathrm{sin}\big(\omega (t-R)\big)\,,\; \nn \\
& r_3=0\,, \nonumber \\
& v_1=-v\;\mathrm{sin}\big(\omega (t-R)\big)\,,\; \\
& v_2=v\;\mathrm{cos}\big(\omega (t-R)\big)\,,\; \nn \\
& v_3=0\,, \nonumber
\end{align}
where $\omega$ is the orbital frequency. In addition, let us suppose
that the mass of the scalar is either sufficiently large or
sufficiently small that variations of $\tilde{\mathcal{G}}$ over an
orbital period can be neglected.  Then $\tilde{\mathcal{G}}$ will
reduce to the massless Brans-Dicke value in the limit of a low mass
scalar \cite{Will:1989sk}, or to the GR value in the limit of a very
massive scalar. With these two approximations made, we perform the
average over one period and obtain the final result for the power
emitted in tensor gravitational waves in the Brans-Dicke theory:
\begin{align}
\dot{E}=-\f{8}{15}\f{\mathcal{G}^2\mu^2m^2v^2}{r^4}(12-6\xi)\,.
\end{align}
Using the relation $(\dot{P}/P)=-\f{3}{2}(\dot{E}/E)$ as well as
the Newtonian result (following from the virial theorem) that
$E=T+V=-\f{1}{2}\mu v^2$ to eliminate $v$, we finally obtain the
fractional period decay due to the emission of tensor gravitational
radiation
\begin{align}
\label{tensor_period_decay}
\f{\dot{P}}{P}=-\f{8}{5}\f{\mathcal{G}^2\mu m^2}{r^4}(12-6\xi)\,.
\end{align}
We stress again that this result is only valid in the limit where
$m_s$ is such that either $e^{-m_sr}\approx 1$, in which case
$\mathcal{G}$ reduces to the massless Brans-Dicke value
\cite{Will:1989sk}, or $e^{-m_sr}\rightarrow 0$, in which case
$\mathcal{G}$ reduces to the GR value.

\subsection{Scalar radiation}

The general expression for the radiated power due to scalar radiation
in Brans-Dicke theory is \cite{Will:1993ns}
\begin{align}
\label{power_loss_scalar}
\dot{E}=-\f{R^2}{32\pi}\phi_0^{-1}(4\omega_{\rm BD}+6)\big\langle\oint\varphi_{,0}\varphi_{,0}d\Omega\big\rangle\,,
\end{align}
where the angular brackets represent the average over one orbital
period.

We can solve Eq.~\eqref{weak_scalar_fe} by using the retarded Green's
function for the massive wave operator $\Box-m_s^2$:
\begin{widetext}
\begin{eqnarray}
G(t-t',\mathbf{R-r'})= \f{\delta(t-t'-|\mathbf{R-r'}|)}{|\mathbf{R-r'}|}
-\Theta(t-t'-|\mathbf{R-r'}|)\f{m_sJ_1(m_s\sqrt{(t-t')^2-|\mathbf{R-r'}|^2})}{\sqrt{(t-t')^2-|\mathbf{R-r'}|^2})}\,,
\end{eqnarray}
\end{widetext}
where $J_1$ is the Bessel function of the first kind, and $\Theta$ is
the Heaviside function (see \cite{morsefeshbach} for a detailed
derivation of this result). Now we can write the general solution to
\eqref{weak_scalar_fe} as
%
$\varphi= \varphi_B+\varphi_m$,
%
where
\begin{widetext}
\begin{eqnarray}
\label{weak_scalar_general_sol}
\varphi_B(t,\mathbf{R}) &=& 4\int\int_\mathscr{N} \f{S(t',\mathbf{r'})\delta(t-t'-|\mathbf{R-r'}|)}{|\mathbf{R-r'}|}d^3\mathbf{r'}dt'\,, \\
\varphi_m(t,\mathbf{R}) &=& -4\int\int_\mathscr{N} \f{m_s S(t',\mathbf{r'}) 
J_1(m_s\sqrt{(t-t')^2-|\mathbf{R-r'}|^2})}{\sqrt{(t-t')^2-|\mathbf{R-r'}|^2}} 
\Theta(t-t'-|\mathbf{R-r'}|)d^3\mathbf{r'}dt' \nonumber \\
&=& -4\int_\mathscr{N} d^3\mathbf{r'}
 \times\int_{0}^\infty\f{J_1(z)S(t-\sqrt{|\mathbf{R-r'}|^2+(\f{z}{m_s})^2},\mathbf{r'})}{\sqrt{|\mathbf{R-r'}|^2+(\f{z}{m_s})^2}}dz\,, \nonumber
\end{eqnarray} 
\end{widetext}
the spatial integration is over the near zone $\mathscr{N}$, and in
the last line we have made the substitution
$z=m_s\sqrt{(t-t')^2-|\mathbf{R-r'}|^2}$.

Taking the field point to be in the radiation zone
($|\mathbf{R}|\gg|\mathbf{r'}|$) and making the slow-motion
approximation, we can expand the $\mathbf{r'}$ dependence of the
integrand and write the general solutions
\eqref{weak_scalar_general_sol} as
\begin{widetext}
\begin{eqnarray} 
\label{weak_scalar_exp_sol_massless}
\varphi_B&=&\f{4}{R}\sum_{m=0}^\infty\f{1}{m!}\f{\partial^m}{\partial t^m}\int_\mathscr{M}d^3\mathbf{r'}S(t-R,\mathbf{r'})(\mathbf{n \cdot r'})^m\,, \\
\label{weak_scalar_exp_sol_massive} 
\varphi_m&=&  
-\f{4}{R}\sum_{m=0}^\infty\f{1}{m!}\f{\partial^m}{\partial t^m}\int_\mathscr{M}d^3\mathbf{r'}(\mathbf{n\cdot r'})^m \times\int_0^\infty dz\f{S(t-\sqrt{R^2+(\f{z}{m_s})^2},\mathbf{r'})J_1(z)}{(1+(\f{z}{m_s R})^2)^\f{m+1}{2}}\,.
\end{eqnarray} 
\end{widetext}
We are now in a position to substitute the post-Newtonian expression
for the source $S$ into \eqref{weak_scalar_exp_sol_massless} and
\eqref{weak_scalar_exp_sol_massive} and obtain an expression for the
gravitational waveform $\varphi(t,\mathbf{R})$ in the far-field,
slow-motion limit. We must first specialize to a two-body system with
the center of mass at the origin, using \eqref{two-body}. Performing
the integration and retaining terms up to order $O(\f{mv^2}{R})$ and
$O(\f{m^2}{Rr'})$ in the monopole ($m=0$) and quadrupole ($m=2$)
terms, and $O(\f{mv}{R})$ in the dipole terms ($m=1$), we obtain
(modulo time-independent terms that are uninteresting, as we
ultimately require $\varphi_{,0}$ in order to calculate the radiated
power)
%
%
\begin{widetext}
\begin{eqnarray}
\label{varphiB} 
\varphi_B&=&2\alpha R^{-1}\mu\Big[\Gamma\mathbf{(n\cdot v)^2}-\frac{1}{2}\Gamma v^2-\tilde{\mathcal{G}}\Gamma m \f{(\mathbf{n\cdot r})^2}{r^3} -(2-\xi)\Gamma'\f{m}{r}-(2\Lambda-(2-\xi)\Gamma')\f{m}{r}e^{-m_sr}-2\mathcal{S}(\mathbf{n\cdot v})\Big]\,, \\
\label{varphiM}
\varphi_m&=& - 2\alpha R^{-1}\mu\Big(\Gamma I_3\big[\mathbf{(n\cdot v)^2}\big]-\frac{1}{2}\Gamma I_1\big[v^2\big]
 -\Gamma I_3\big[\tilde{\mathcal{G}} m \f{(\mathbf{n\cdot r})^2}{r^3}\big]-(2-\xi)\Gamma'I_1\big[\f{m}{r}\big] \nonumber \\
&-&(2\Lambda-(2-\xi)\Gamma')I_1\big[\f{m}{r}e^{-m_sr}\big]-2\mathcal{S}I_2\big[(\mathbf{n\cdot v})\big]\Big)\,.
\end{eqnarray} 
\end{widetext}
Here we have defined 
\begin{align}
\label{gamma_definitions}
& \Gamma\equiv 1-2\f{s_1m_2+m_1s_2}{m}\,, \nonumber \\
& \Gamma'\equiv 1-s_1-s_2\,, \nonumber \\
& \Lambda\equiv \mathcal{G}\Gamma'-\xi\big((1-2s_1)s_2'+(1-2s_2)s_1'\big)\,,
\end{align}
%
%
and the terms $I_n\big[f(t)\big]$ represent the integrals
\begin{align}
\label{I_def}
I_n\big[f(t)\big]=\int_0^\infty\f{f(t-\sqrt{R^2+(\f{z}{m_s})^2})J_1(z)}{(1+(\f{z}{m_s R})^2)^{\f{n}{2}}}dz\,,
\end{align}
where the integration over $z$ has yet to be performed, and it is understood that the time-dependent terms 
in \eqref{varphiB} and \eqref{varphiM} (replacing $f(t)$ in \eqref{I_def}) are the components of 
$\mathbf{r}$ and $\mathbf{v}$.
%
%
As in the calculation of the tensor component, we assume that
$\tilde{\mathcal{G}}$ is approximately constant over an orbital period
($\tilde{\mathcal{G}} \rightarrow \mathcal{G}$), and we specialize to
a circular orbit parametrized by \eqref{circle}. Taking the partial
time derivative of $\varphi$ we find
\begin{align}
\varphi_{,0}=&2\alpha R^{-1}\mathcal{G}m\mu\Big[2\mathcal{S}\Big(\f{n^i r^i}{r^3}-I_2\big[\f{n^i r^i}{r^3}\big]\Big) \nonumber \\
&-4\Gamma\Big(\f{n^i n^j v^i r^j}{r^3}-I_3\big[\f{n^i n^j v^i r^j}{r^3}\big]\Big)\Big],
\end{align}
where we have used \eqref{eom} to replace $\ddot{r}^i$ (to leading
order) where necessary. The first and second terms represent the
dipole and quadrupole contributions respectively; note that there is
no monopole contribution to leading order in the circular orbit case.
Substituting this into \eqref{power_loss_scalar} and performing the
integration over the solid angle via the identity
\eqref{unit_int_identity}, we obtain
\begin{widetext}
\begin{eqnarray} 
\dot{E}&=&-\mathcal{G}^2m^2\mu^2\xi\Big\langle\f{2}{3}\mathcal{S}^2\Big(\f{r^i}{r^3}-I_2\big[\f{r^i}{r^3}\big]\Big)\Big(\f{r^i}{r^3}-I_2\big[\f{r^i}{r^3}\big]\Big) 
+\f{8}{15}\Gamma^2\Big(\f{r^iv^j}{r^3}-I_3\big[\f{r^iv^j}{r^3}\big]\Big)\Big(\f{r^iv^j}{r^3}-I_3\big[\f{r^iv^j}{r^3}\big]\Big) \\
&+&\f{8}{15}\Gamma^2\Big(\f{r^iv^j}{r^3}-I_3\big[\f{r^iv^j}{r^3}\big]\Big)\Big(\f{r^jv^i}{r^3}-I_3\big[\f{r^jv^i}{r^3}\big]\Big)\Big\rangle \nonumber \\
&=&-\f{\mathcal{G}^2\mu^2m^2\xi}{r^4}\left[\f{2}{3}\mathcal{S}^2\big(1-2Z_2(R;m_s,\omega)+W_2(R;m_s,\omega)\big)
+\f{8}{15}\Gamma^2v^2\big(1-2Z_3(R;m_s,2\omega)+W_3(R;m_s,2\omega)\big)\right]\,, \nonumber
\end{eqnarray} 
\end{widetext}
where in the second line we have performed the average over one
orbital period and we have defined
\begin{widetext}
\begin{eqnarray}
\label{nasty_integrals} 
Z_n(R;m_s,\omega)&\equiv&\mathrm{cos}(\omega R)C_n(R;m_s,\omega)+\mathrm{sin}(\omega R)S_n(R;m_s,\omega)\,, \nonumber \\
W_n(R;m_s,\omega)&\equiv& |C_n(R;m_s,\omega)|^2+|S_n(R;m_s,\omega)|^2\,, \nonumber \\
C_n(R;m_s,\omega)&\equiv&\int_0^\infty\mathrm{cos}\Big(\omega R\sqrt{1+(\f{z}{m_sR})^2}\Big)\f{J_1(z)}{\big(1+(\f{z}{m_sR})^2\big)^\f{n}{2}}dz\,, \nonumber \\
S_n(R;m_s,\omega)&\equiv&\int_0^\infty\mathrm{sin}\Big(\omega R\sqrt{1+(\f{z}{m_sR})^2}\Big)\f{J_1(z)}{\big(1+(\f{z}{m_sR})^2\big)^\f{n}{2}}dz\,.
\end{eqnarray} 
\end{widetext}

To get the total power radiated we must perform the integrals in the
limit $R\rightarrow\infty$ in which they have closed form
solutions. The evaluation of these integrals is discussed in Appendix
\ref{sec:integration_appendix}. Performing the integrals, we obtain
\begin{align}
\dot{E}=&-\f{\mathcal{G}^2m^2\mu^2\xi}{r^4}\Big[\f{8}{15}\Gamma^2v^2\Big(\f{4\omega^2-m_s^2}{4\omega^2}\Big)^2\Theta(2\omega-m_s) \nonumber \\
&+\f{2}{3}\mathcal{S}^2\f{\omega^2-m_s^2}{\omega^2}\Theta(\omega-m_s)\Big].
\end{align}
Using again $(\dot{P}/P)=-\f{3}{2}(\dot{E}/E)$, and
$E=-\f{1}{2}\f{\mathcal{G}m\mu}{r}=-\f{1}{2}\mu v^2$, we can eliminate $v$ and
find for the fractional period derivative due to scalar radiation:
\begin{align}
\label{scalar_power_emitted}
\f{\dot{P}}{P}=& -\f{96}{5}\f{\mathcal{G}^2\mu m^2}{r^4}\f{\Gamma^2}{12}\xi\Big(\f{4\omega^2-m_s^2}{4\omega^2}\Big)^2\Theta(2\omega-m_s) \nonumber \\
&-\f{2\mathcal{G}\mu m}{r^3}\mathcal{S}^2\xi\f{\omega^2-m_s^2}{\omega^2}\Theta(\omega-m_s)\,.
\end{align}
Combining this with the result for the tensor gravitational radiation
contribution \eqref{tensor_period_decay}, we finally obtain the result
quoted in Eq.~\eqref{period_decay_mbd} of the introduction.

\section{Obtaining bounds on $(\omega_{\rm BD},m_s)$}\label{sec:bounds}

In this section we compare our results for the period derivative of
compact binaries, the Shapiro delay and the Nordtvedt effect against
recent observational data to draw exclusion plots in the
two-dimensional parameter space of the theory, $(m_s,\,\omega_{\rm
  BD})$. Figure \ref{fig:bounds_all} in the introduction summarizes
our main results.

%
%

\subsection{Bounds from $\dot{P}$ in compact binaries}
\label{sec:Pdot_bounds}

Due to the presence of the difference in sensitivities ($\mathcal{S}=s_1-s_2$)
in the dipole contribution to the period decay
\eqref{period_decay_mbd}, the best candidate systems for drawing
exclusion plots in the $(\omega_{\rm BD},m_s)$ plane are mixed
binaries. White dwarf-neutron star (WD-NS) binaries are particularly
suitable due to the large difference in sensitivities ($\sim10^{-4}$
and $\sim0.2$ for WDs and NSs, respectively \cite{Zaglauer:1992bp}).
To our knowledge, there are three such systems for which accurate
measurements of $\dot{P}$ and the other necessary parameters have been
made (to date): PSRs J0751+1807, J1012+5307, J1141-6545. A summary of
the observations and the relevant references are provided in Appendix
\ref{sec:binary_appendix}. 

In principle, there are two more systems that are of interest to our
current purposes: PSR J1738+0333 \cite{2010arXiv1006.0642F} and PSR
J1802-2124 \cite{Ferdman:2010rk}. Both of these systems have very
small eccentricities, which means that the result derived in this
paper can be used ``out of the box'', with no need to generalize our
calculation to eccentric binaries. In the case of PSR J1738+0333, a
relatively accurate measurement of $\dot{P}$ has been achieved (with
error $\sim 30\%$), but the masses of the component stars have yet to
be determined \cite{2010arXiv1006.0642F}. PSR J1802-2124 is in
precisely the opposite situation: the masses of the components have
been measured to reasonable precision, but a precise measurement of
$\dot{P}$ has yet to be achieved. This is anticipated in the near
future \cite{Ferdman:2010rk}.


The general approach to obtaining bounds on $(\omega_{\rm BD},m_s)$ using
observations of the period derivative of mixed binaries is as
follows. Firstly, we need to write \eqref{period_decay_mbd} in
terms of the observables relevant to the system under inspection. For 
circular binaries, the relevant observables are the stellar 
masses (including the mass ratio $q$) and the period. Recasting
\eqref{period_decay_mbd} into these observables we obtain
\begin{align}
\dot{P}=\dot{P}_\mathrm{GR}\left[\mathcal{G}^{-\f{4}{3}}\f{\kappa_1}{12}+\f{5}{96}m^{-\f{2}{3}}\Big(\f{2\pi}{P}\Big)^{-\f{2}{3}}2\mathcal{S}^2\kappa_D\right]\,,
\end{align}
where $\dot{P}_\mathrm{GR}$ is the prediction from GR, given by
\begin{align}
\label{pdotgrcirc}
\dot{P}_\mathrm{GR}=-\f{192\pi}{5}\f{q\;m^\f{5}{3}}{(1+q)^2}\left(\f{2\pi}{P}\right)^\f{5}{3}.
\end{align}
For mildly eccentric binaries, provided the eccentricity is small
enough, we can get approximate bounds using the results obtained here
for the circular case. In these instances, we can use the measured
periastron shift $\dot{\omega}$, period, and the mass ratio $q$ of the
binary. Recasting \eqref{period_decay_mbd} in terms of these
observables (using the results for the periastron advance quoted in
section \ref{sec:periastron} and Kepler's third law to eliminate $m$
and $r$) we obtain
\begin{align}
\dot{P}=\dot{P}_\mathrm{GR}\left[\f{\mathcal{G}^2}{\mathcal{P}^{-\f{5}{2}}}\f{\kappa_1}{12}+\f{5}{16}\f{2\pi}{P\dot{\omega}}\mathcal{S}^2\f{\kappa_D}{2}\right]\,,
\end{align}
where
\begin{align}
\label{pdotgrecc}
\dot{P}_\mathrm{GR}=-\f{4 q P}{(1+q)^2}\f{8}{15\sqrt{3}}\left(\f{P}{2\pi}\right)^\f{3}{2}\dot{\omega}^\f{5}{2}\,.
\end{align}
With the predicted $\dot{P}$ written in terms of the relevant set of parameters,
we are in a position to compare it to observations; the predicted $\dot{P}$
and observed $\dot{P}_\mathrm{obs}$ are consistent to $n\sigma$ confidence provided
that
\begin{align}
\label{set}
|\dot{P}_\mathrm{obs}-\dot{P}(\xi,m_s)|\leq n\sigma,
\end{align}
where $\sigma$ is the combined uncertainty of $\dot{P}_\mathrm{obs}$
and the predicted $\dot{P}$, and where we should remember that the
latter is uncertain due to uncertainties in the observables (such as
stellar masses and period). In order to obtain an upper bound on $\xi$
(and hence a lower bound on $\omega_{\rm BD}$) to 95\% confidence for
a range of scalar masses, in Figure \ref{fig:bounds_all} we simply
plot the contour in the $(\omega_{\rm BD},m_s)$-plane associated with
$|\dot{P}_\mathrm{obs}-\dot{P}(\xi,m_s)|=2\sigma$.

\subsubsection{Bounds from neutron star-neutron star binaries}

The presence of dipole radiation in mixed binaries suggests that these
should be the best candidates for obtaining the most stringent bounds
on $(\omega_{\rm BD},m_s)$. However, it is worth looking into the
bounds that could be obtained from observations of neutron
star-neutron star (NS-NS) binaries. Since the sensitivities of the two
component stars are nearly identical in this case ($\mathcal{S}\simeq
0$: cf. \cite{Zaglauer:1992bp}), the expression for the period
derivative reduces to
\begin{align}
\dot{P}=\dot{P}_\mathrm{GR}\f{\kappa_1}{12}\,.
\end{align}
Expanding to linear order in $\xi$ we can write this as
\begin{align}
\dot{P}=\dot{P}_\mathrm{GR}\left(1+\xi\chi\right)\,, \nn \\
\end{align}
where
\begin{align}
\chi=&\f{\Gamma^2}{12}\left(\f{4\omega^2-m_s^2}{4\omega^2}\right)^2\Theta(2\omega-m_s)-\f{5}{6} \nn \\
&+\f{1}{3}(1-2s_1)(1-2s_2)\,.
\end{align}

We can also write the observed period derivative as
\begin{align}
\dot{P}_\mathrm{obs}=\dot{P}_\mathrm{GR}(1+\delta)\,,
\end{align}
where $\delta$ is the fractional deviation of the observed value from
the GR prediction.  Applying the condition \eqref{set} to the above
two equations we find that the predicted $\dot{P}$ is consistent with
the observed $\dot{P}_\mathrm{obs}$ to $2\sigma$ confidence provided
that
\begin{align}
|\xi\chi-\delta|\leq 2\sigma\,,
\end{align}
where $\sigma$ is now the combined uncertainty of $\chi$ and the
observed deviation from GR, $\delta$. Since the correction $\chi$ in
the above is of order unity, we conclude that bounds competitive with
the most stringent bounds obtained here (from the Cassini Shapiro
delay measurements, and those that are expected from a rigorous
analysis of PSR J1141-6545) could only be obtained from a $\dot{P}$
measurement to a precision of $\sim 0.01\%$. Since the current best
measurements of $\dot{P}$ for NS-NS binaries are not yet close to this
precision, we conclude that the bounds that would be obtained by
analyzing such systems would be significantly weaker than the most
stringent bounds obtained here. For NS-NS systems for which $\dot{P}$
has been measured to a precision of $\sim 1\%$ (such as PSR J0737-3039
\cite{Kramer:2006nb}), we would expect to obtain relatively weak
bounds, comparable to those obtained here from the quasi-circular
WD-NS binary PSR J1012+5307.

\subsection{Bounds from Cassini time-delay data}
\label{sec:shapiro_bounds}

The Shapiro delay has been measured in the Solar System to remarkable
precision by radio tracking of the Cassini spacecraft in 2002
\cite{Bertotti:2003rm}. In theories containing only massless fields,
these observations are tantamount to a measurement of the PPN
parameter $\gamma$. 
This has been measured to be
\begin{align}
\label{gammacassini}
\gamma^{\rm Cassini}=1+(2.1\pm2.3)\times10^{-5}=1+\delta\pm\epsilon.
\end{align}
As discussed in section \ref{sec:shapiro_delay}, the mass of the
scalar in the massive Brans-Dicke theory prohibits us from using the
PPN formalism in the conventional manner, and the concept of constant
PPN parameters breaks down (see also \cite{Perivolaropoulos:2009ak}).
In section \ref{sec:shapiro_delay} we derived an expression for the Shapiro
delay in the massive Brans-Dicke theory, and we defined a quantity $\tilde{\gamma}$
which is analogous to the PPN parameter $\gamma$ (at least in the context of Shapiro delay)
and can be directly compared to the measured value of $\gamma$ to obtain an exclusion
region in the $(m_s,\,\omega_{\rm BD})$ plane. Comparing the derived expression for
$\tilde{\gamma}$ \eqref{gammatilde} with the Cassini
measurement of $\gamma$ \eqref{gammacassini}, we require that 

%
\begin{align}
\alpha<e^{m_sr}\f{2\epsilon-\delta}{(2+\delta-2\epsilon)}
\end{align}
to $95\%$ confidence. The resulting bounds on $\xi$ and
$\omega_{\rm BD}$ are plotted in Figure \ref{fig:bounds_all} by solid
black lines (cf. also \cite{Perivolaropoulos:2009ak}).

We find that $\omega_{\rm BD}>40000$ for a range of scalar masses
$m_s<2.5\times10^{-20}\mathrm{eV}$, to 95\% confidence. This is around
one order of magnitude more stringent than the bounds provided by the
observations of gravitational radiation damping in binary systems. In
the limit $1/m_s\ll 1\mathrm{AU}$, $\omega_{\rm BD}$ can take on any value
(as long as $\omega_{\rm BD} >-3/2$).  
\subsection{Bounds from Lunar Laser Ranging observations}
\label{sec:nordtvedt_bounds}

The most precise measurement of the Nordtvedt effect to date comes
from the Lunar Laser Ranging experiment \cite{2010A&A...522L...5H}
\begin{align}
\eta_\mathrm{N}^{LLR}=(−0.6 \pm 5.2)\times10^{-4}=\delta\pm\epsilon\,.
\end{align}
Comparing this observed value to Eq.~\eqref{nordtvedt_effective} and
neglecting the small sensitivity of the Sun, we require that
\begin{align}
|\xi(1+m_sr)e^{-m_sr}-\delta|\leq2\epsilon
\end{align}
to 95\% confidence, from which we obtain exclusion regions in the
$(m_s,\,\omega_{\rm BD})$ plane. These are displayed in Figure
\ref{fig:bounds_all} by dotted blue lines.

\section{Conclusions}

In this paper we set constraints on massive Brans-Dicke (or
Bergmann-Wagoner) theories with an action of the form
(\ref{action_general}) with $\omega(\phi)=\omega_{\rm BD}$, assuming
that only a mass term is present in the expansion of $M(\phi)$ around
some cosmologically imposed value $\phi_0$.  In particular we computed
the orbital period derivative for quasicircular binaries. From an
observational standpoint it will be important to generalize our work
to eccentric (and possibly spinning) binaries, that could yield more
stringent constraints on scalar-tensor theories. It will also be
interesting to explore possible bounds on massive scalar-tensor
theories that could result from Earth- and space-based
gravitational-wave observations of compact binaries, along the line
of Refs.~\cite{Will:1994fb,Berti:2004bd,Berti:2005qd}.

A second obvious generalization of our work will consist in relaxing
our assumptions on the form of $\omega(\phi)$ and $M(\phi)$. More
generic assumptions on these functions are necessary for a deeper
understanding of binary dynamics in the context of modified gravity
models that try to explain cosmological observations. It will be
interesting to verify whether time-varying boundary conditions on the
scalar field may lead to interesting binary dynamics
\cite{Horbatsch:2011ye}.

Last but not least, full numerical relativity simulations of compact
binaries in scalar-tensor theories are under investigation by several
groups (see e.g. \cite{Salgado:2008xh,Healy:2011ef}). Numerical
progress in evolving binary dynamics in alternative theories is
important, as it could reveal strong-field effects that may be
inaccessible to post-Newtonian or perturbative calculations.

\noindent
\begin{acknowledgments}
  We are grateful to Vitor Cardoso, Yanbei Chen, Samaya Nissanke,
  Ulrich Sperhake, Michele Vallisneri and Helvi Witek for
  discussions. We are particularly grateful to Leonardo Gualtieri,
  Michael Horbatsch and Paolo Pani for suggestions and detailed
  comments on the manuscript, and to Michael Horbatsch for checking
  that our dipolar and quadrupolar fluxes match Eq.~(6.40) in
  \cite{Damour:1992we} when $m_s\to 0$. J.A. was supported by a LIGO
  SURF Fellowship at Caltech.  E.B. was supported by NSF Grant
  PHY-0900735 and by NSF CAREER Grant PHY-1055103. C.M.W. was
  supported by NSF Grant PHY-0965133. C.M.W. is grateful for the
  hospitality of the Institut d'Astrophysique de Paris, where part of
  this work was carried out.
\end{acknowledgments}

\appendix

\section{Post-Newtonian expansion of the scalar field and of the metric}
\label{sec:pn_appendix}

Here we provide details of the derivation of the post-Newtonian
expansions \eqref{pn_phi}, \eqref{pn_g00}, \eqref{pn_g0i} and
\eqref{pn_gij}. We follow very closely the method outlined in
\cite{Will:1993ns}. For our current purposes, we must solve the field
equations \eqref{mbd_tensor_fe} and \eqref{mbd_scalar_fe} to the
following orders:
\begin{align}
&\varphi\sim O(2) + O(4), \nn \\
&h_{00}\sim O(2)+O(4), \nn \\
&h_{0i}\sim O(3), \nn \\
&h_{ij}\sim O(2).
\end{align}
We will do this in a number of steps, as described in the following.
\subsection{Step 1: $\varphi$ to order $O(2)$}
To the lowest order, the scalar field equation \eqref{mbd_scalar_fe}
reduces to
\begin{align}
\label{mbd_scalar_fe_lowest_order}
\left(\nabla^2-m_s^2\right)\f{\varphi}{\phi_0}=8\pi\alpha\phi_0^{-1}T^*.
\end{align}
Expanding the modified stress-energy tensor $T^*$ to lowest order we
obtain
\begin{align}
\label{set_O2}
T^* &= \sum_a m_a(2s_a-1)\delta^3(\mathbf{x}-\mathbf{x}_a) \nn \\
&= -\f{1}{4\pi}\sum_a m_a(2s_a-1)\left(\nabla^2-m_s^2\right)\f{e^{-m_sr_a}}{r_a}.
\end{align}
Substituting this into \eqref{mbd_scalar_fe_lowest_order}, we find the
solution for $\varphi$ to $O(2)$
\begin{align}
\f{\varphi}{\phi_0}=\xi\sum_a\f{m_a}{r_a}e^{-m_sr_a}\left(1-2s_a\right).
\end{align}
\subsection{Step 2: $h_{00}$ to $O(2)$}
The $00$-component of the tensor field equation \eqref{mbd_tensor_fe}
to O(2) is given by
\begin{align}
\label{tfeO2}
R_{00}=-\f{1}{2}\nabla^2h_{00}=8\pi
\left(T_{00}+\f{T}{2}\right)-
\f{1}{2}\nabla^2\left(\f{\varphi}{\phi_0}\right).
\end{align}
In a similar fashion to \eqref{set_O2}, we write for the stress-energy
tensor (to lowest order)
\begin{align}
\label{setO2a}
T &= -T_{00} = -\sum_a m_a\delta^3(\mathbf{x}-\mathbf{x}_a) \nn \\
& = \f{1}{4\pi}\sum_a m_a\nabla^2\left(\f{1}{r_a}\right).
\end{align}
Substituting this along with the derived $O(2)$ expression for
$\varphi$ into \eqref{tfeO2}, we obtain the $O(2)$ solution for
$h_{00}$:
\begin{align}
h_{00}=2\phi_0^{-1}\sum_a\f{m_a}{r_a}
\left[1+\alpha\left(1-2s_a\right)e^{-m_sr_a}\right].
\end{align}
\subsection{Step 3: $h_{ij}$ to $O(2)$}
The $ij$-component of the Ricci tensor to O(2) is given by
\begin{align}
R_{ij} = -\f{1}{2}(\nabla^2h_{ij}-h_{00,ij}+h^k_{k,ij}-h^k_{i,kj}-h^k_{j,ki}).
\end{align}
Imposing the gauge condition
\begin{align}
\label{gauge1}
h^\mu_{i,\mu}-\f{1}{2}h^\mu_{\mu,i}=\left(\f{\varphi}{\phi_0}\right)_{,i},
\end{align}
we can write the $ij$-component of tensor field equation
\eqref{mbd_tensor_fe} to O(2) as
\begin{align}
\nabla^2h_{ij}=8\pi\phi_0^{-1}T\delta_{ij} + 
\delta_{ij}\nabla^2\left(\f{\varphi}{\phi_0}\right).
\end{align}
Using Eq.~\eqref{setO2a} for the stress-energy tensor and substituting
the derived O(2) expression for $\varphi$ into the expression above,
we obtain the solution for $h_{ij}$ to $O(2)$:
\begin{align}
h_{ij}=\delta_{ij}2\phi_0^{-1}
\sum_a\f{m_a}{r_a}\left[1-\alpha(1-2s_a)e^{-m_sr_a}\right].
\end{align}
\subsection{Step 4: $h_{0i}$ to $O(3)$}
The $0i$-component of the Ricci tensor to O(3) is given by
\begin{align}
R_{0i}=-\f{1}{2}\left(\nabla^2h_{0i}
-h^{\;\;\;,k}_{0k\;\;,i}+h^{k}_{\;k,0i}-h^{\;\;\;,k}_{ki\;\;,0}\right).
\end{align}
Imposing the further gauge condition
\begin{align}
\label{gauge2}
h^\mu_{0,\mu}-\f{1}{2}h^\mu_{\mu,0}=
-\f{1}{2}h_{00,0}+\left(\f{\varphi}{\phi_0}\right)_{,0}
\end{align}
this reduces to
\begin{align}
R_{0i}=-\f{1}{2}\nabla^2h_{0i}+
\f{1}{2}\left(\f{\varphi}{\phi_0}\right)_{,0i}-\f{1}{12}h^{k}_{\;k,0i}.
\end{align}
We can hence write the $0i$-component of the tensor field equation
\eqref{mbd_tensor_fe} to $O(3)$ as
\begin{align}
\label{h0ieqn}
-\f{1}{2}\nabla^2h_{0i} = 
8\pi\phi_0^{-1}T_{0i}+
\f{1}{2}\left(\f{\varphi}{\phi_0}\right)_{,0i}+\f{1}{12}h^k_{\;k,0i}.
\end{align}
The $0i$-component of the stress-energy tensor to lowest order is
given by
\begin{align}
T_0^i= & -\sum_a m_av_a^i\delta^3(\mathbf{x}-\mathbf{x}_a) \nn \\
&= \f{1}{4\pi}\sum_a m_a v_a^i\nabla^2\left(\f{1}{r_a}\right).
\end{align}
In order to write the $\varphi_{,00}$ term in the form $\nabla^2\chi$,
we must find a particular solution to $\nabla^2\chi =
e^{-m_sr}/r$. Taking care to ensure that the chosen solution $\chi$ is
such that the correct limit is obtained as $m_s\rightarrow 0$, we
write
\begin{align}
\label{gf3}
&\nabla^2\left(\f{e^{-m_sr_a}+m_sr_a-1}{m_s^2r_a}\right)=\f{e^{-m_sr_a}}{r_a},
\end{align}
Noting also that $\nabla^2(r_a/2)=r_a$, we can re-write the second and
third terms in \eqref{h0ieqn} as
\begin{align}
& h^k_{k,0i}=\f{6}{\phi_0}
\sum_a m_a\partial_i\partial_0
\left[\f{r_a}{2}-\alpha(1-2s_a)\f{e^{-m_s r_a}+m_sr_a -1}{m_s^2r_a}\right], \nn \\
& \left(\f{\varphi}{\phi_0}\right)_{,0i}=
\xi\sum_a m_a(1-2s_a)\partial_i\partial_0
\left(\f{e^{-m_s r_a}+m_sr_a -1}{m_s^2r_a}\right).
\end{align}
Substituting these into \eqref{h0ieqn}, we obtain the solution for
$h_{0i}$ to $O(3)$ given in Eq.~\eqref{pn_g0i}.
\subsection{Step 5: $\varphi$ to $O(4)$}
Expanding $\Box_g \varphi$ to $O(4)$ and recalling the definition of
$\theta^{\mu\nu}$ in Eq.~\eqref{perturb}, we obtain
\begin{align}
\Box_g\phi=
\left(1+\f{1}{2}\theta+\f{\varphi}{\phi_0}\right)
\Box_\eta\f{\varphi}{\phi_0}-
\theta^{\mu\nu}\f{\varphi_{,\mu\nu}}{\phi_0}-
\f{\varphi_{,\alpha}\varphi^{,\alpha}}{\phi_0^2}.
\end{align}
The scalar field equation \eqref{mbd_scalar_fe} to $O(4)$ can hence be
written as
\begin{align}
\label{scalO4}
\left(\nabla^2-m_s^2\right)\f{\varphi}{\phi_0} =& 8
\pi\alpha\phi_0^{-1}T^*\left(1-\f{1}{2}\theta-\f{\varphi}{\phi_0}\right) \nn \\ 
& + \left(\f{\varphi}{\phi_0}\right)_{,00} + 
\left(\nabla\f{\varphi}{\phi_0}\right)^2.
\end{align}
Expanding the modified stress-energy tensor to the required order we
find
\begin{align}
T^* = & \sum_a m_a\Big[(2s_a - 1) + \f{1}{2}(1-2s_a)v_a^2 \nn \\
& - \f{3}{4}(1-2s_a)\theta - \left(2s_a'-2s_a^2+\f{3}{2}\right)
\f{\varphi}{\phi_0}\Big].
\end{align}
In a similar fashion to step 4, we wish to write the $\varphi_{,00}$
term in the form $(\nabla^2-m_s^2)\chi$, and we require a particular
solution to $(\nabla^2-m_s^2)\chi=e^{-m_sr}/r$ such that the correct
limit is obtained as $m_s\rightarrow 0$. To this end we write
\begin{align}
\label{gf2}
(\nabla^2-m_s^2)\left(\f{1-e^{-m_sr}}{2m_s}\right)=\f{e^{-m_sr}}{r},
\end{align}
and hence write the second term on the right hand side of
Eq.~\eqref{scalO4} as
\begin{align}
\left(\f{\varphi}{\phi_0}\right)_{,00}=
(\nabla^2-m_s^2)\xi\sum_a m_a(1-2s_a)
\partial_0\partial_0\left(\f{1-e^{-m_sr}}{2m_s}\right).
\end{align}
The third term on the right hand side of Eq.~\eqref{scalO4} can be
re-written (to the required order) as
\begin{align}
\left(\nabla\f{\varphi}{\phi_0}\right)^2 =& 
\f{1}{2}(\nabla^2-m_s^2)\left(\f{\varphi}{\phi_0}\right)^2 \nn \\
&-\left(\f{\varphi}{\phi_0}\right)(\nabla^2-m_s^2)
\left(\f{\varphi}{\phi_0}\right).
\end{align}
Substituting the above along with the derived $O(2)$ expressions for
$\varphi$, $h_{00}$ and $h_{ij}$ into \eqref{scalO4}, we obtain the
result given in Eq.~\eqref{pn_phi}.
\subsection{Step 6: $h_{00}$ to $O(4)$}
The $00$-component of the tensor field equation \eqref{mbd_tensor_fe}
to $O(4)$ is given by
\begin{align}
\label{tfeO4}
R_{00}=&-\f{1}{2}\nabla^2h_{00} +
\left(\f{\varphi}{\phi_0}\right)_{,00} + \f{1}{2}\nabla h_{00} 
\nabla\f{\varphi}{\phi_0} \nn \\
&- \f{1}{2}(\nabla h_{00})^2+\f{1}{2}h_{00}\nabla^2h_{00} - 
\f{\varphi}{\phi_0}\nabla^2h_{00} \nn \\
&= 8\pi\phi_0^{-1}\left(1-\f{\varphi}{\phi_0}\right)
(T_{00}-\f{1}{2}g_{00}T) \nn \\
& + \left(1-\f{\varphi}{\phi_0}\right)
\left(\left(\f{\varphi}{\phi_0}\right)_{,00}+
\f{1}{2}g_{00}\Box_g\left(\f{\varphi}{\phi_0}\right)\right),
\end{align}
where we have used the gauge conditions \eqref{gauge1} and
\eqref{gauge2} to reduce the expression for $R_{00}$ into a convenient
form.  The term involving the stress-energy tensor on the right hand
side is given (to the required order) by
\begin{align}
T_{00}-\f{1}{2}g_{00}T =& 
\f{1}{2}\sum_a m_a\delta^3(\mathbf{x}-\mathbf{x}_a)
\Big[1+\f{3}{4}v_a^2+\f{5}{4}\theta \nn \\
&+\f{1}{2}(2s_a+1)\left(\f{\varphi}{\phi_0}\right)\Big].
\end{align}
Using this along with
\begin{align}
& (\nabla^2h_{00})^2=\f{1}{2}\nabla^2h^2_{00}-h_{00}\nabla^2h_{00}, \nn \\
& (\nabla^2\f{\varphi}{\phi_0})^2=\f{1}{2}
\nabla^2\left(\f{\varphi}{\phi_0}\right)^2-\f{\varphi}{\phi_0}
\nabla^2\f{\varphi}{\phi_0},
\end{align}
Eq.~\eqref{tfeO4} can be re-written as
\begin{align}
& 2\phi_0^{-1}\sum_a m_a \nabla^2\left(\f{1}{r_a}\right)
\left[1+\f{3}{2}v_a^2+\f{5}{4}\theta+(s_a-\f{1}{2})
\f{\varphi}{\phi_0}\right] \nn \\
& +\nabla^2\f{\varphi}{\phi_0} + (\f{1}{2}\theta - h_{00})\nabla^2
\f{\varphi}{\phi_0} - \nabla^2\left(\f{\varphi}{\phi_0}\right)^2 \nn \\
& - \f{1}{2}\nabla^2h_{00}^2 + 2h_{00}\nabla^2h_{00}-2\f{\varphi}{\phi_0}
\nabla^2h_{00} - \f{1}{2}\left(\f{\varphi}{\phi_0}\right)_{,00} \nn \\
& +2\f{\varphi}{\phi_0}\nabla^2\f{\varphi}{\phi_0} = \nabla^2h_{00}
\end{align}
Using Eq.~\eqref{gf3} to re-write the term involving $\varphi_{,00}$
in the form $\nabla^2\chi$, and substituting in the derived $O(2)$
expressions for $h_{00}$, $h_{ij}$ and $\varphi$, and the O(4)
expression for $\varphi$, we obtain the result presented in
Eq.~\eqref{pn_g00}. Note that there are two contributions to the term
involving the second time derivative in Eq.~\eqref{pn_g00}: one
contribution from $\varphi$ (to $O(4)$) and one from $\varphi_{,00}$.

\section{Evaluation of integrals $C_n(R;m_s,\omega)$ and $S_n(R;m_s,\omega)$ arising in the derivation of the period derivative due to scalar radiation}
\label{sec:integration_appendix}

In reaching the final expression for the power emitted in scalar
gravitational radiation \eqref{scalar_power_emitted}, we were required
to evaluate the integrals $C_n(R;m_s,\omega)$ and $S_n(R;m_s,\omega)$
defined in \eqref{nasty_integrals}.  In this appendix we give details
of the evaluation of these integrals.

Since we are interested in the gravitational radiation in the far
zone, we only need to determine the asymptotic behavior of these
integrals for $R\rightarrow\infty$. Substituting
$u=\sqrt{1+(z/m_sR)^2}$ into \eqref{nasty_integrals} we obtain
\begin{align}
& C_n(R;m_s,\omega)=m_sR\int_1^\infty du\f{\mathrm{cos}(\omega Ru)}{u^{n-1}}\f{J_1(m_sR\sqrt{u^2-1})}{\sqrt{u^2-1}}\,, \nn \\
& S_n(R;m_s,\omega)=m_sR\int_1^\infty du\f{\mathrm{sin}(\omega Ru)}{u^{n-1}}\f{J_1(m_sR\sqrt{u^2-1})}{\sqrt{u^2-1}}\,.
\end{align}
We will discuss the evaluation of $C_n$ only, as the evaluation of
$S_n$ proceeds in exactly the same way. To begin with, let us choose
some $\epsilon$ such that $m_sR\epsilon\gg1$ while $\omega
R\epsilon^2\ll1$ and split up the integral into an integration from $1$
to $1+\epsilon^2/2$ and from $1+\epsilon^2/2$ to $\infty$. In the
first integral, as the argument of the cosine is nearly constant we
can approximate
\begin{align}
\label{boundary_term}
m_sR\int_1^{1+\epsilon^2/2}du\f{\mathrm{cos}(\omega Ru)}{u^{n-1}}&\f{J_1(m_sR\sqrt{u^2-1})}{\sqrt{u^2-1}} \nn \\
&\approx\mathrm{cos}(\omega R)\big(1-J_0(m_sR\epsilon)\big),
\end{align}
with the zeroth order Bessel function $J_0$ given by its asymptotic
value
\begin{align}
J_0(m_sR\epsilon)\sim\sqrt{\f{2}{\pi}}\f{\mathrm{cos}(m_sR\epsilon-\pi/4)}{\sqrt{m_sR\epsilon}}.
\end{align}
For the second integral, we can approximate the Bessel function $J_1$
by its asymptotic value
\begin{align}
J_1(x)\sim\sqrt{\f{2}{\pi}}\f{\mathrm{cos}(x-3\pi/4)}{\sqrt{x}},
\end{align}
and hence the integral can be approximated by
\begin{align}
\sqrt{\f{2}{\pi}}\sqrt{m_sR}\int_{1+\epsilon^2/2}^\infty du\f{\mathrm{cos}(\omega Ru)}{u^{n-1}}\f{\mathrm{cos}(m_sR\sqrt{u^2-1}-3\pi/4)}{\sqrt[4]{u^2-1}^3}.
\end{align}
Performing an integration by parts exactly cancels the corresponding
boundary term in \eqref{boundary_term}; this is not surprising, since
we expect that the result should not depend on the value of
$\epsilon$. In analyzing the above integral, then, we can neglect all
terms arising from the lower endpoint (since a full analysis will show
that they will exactly cancel the terms arising from the upper
endpoint in \eqref{boundary_term}). We are interested in the leading
asymptotic behavior of the above integral; we hence require the
asymptotic behavior of integrals of the type
\begin{align}
I=\f{1}{4}\sqrt{\f{2}{\pi}}\sqrt{m_sR}\int_{1+\epsilon^2/2}^\infty\f{du}{u^{n-1}\sqrt[4]{u^2-1}^3}e^{\rho(u)},
\end{align}
where
\begin{align}
\rho(u)=iR(n_1\omega u+n_2m_s\sqrt{u^2-1})-in_23\pi/4,
\end{align}
with $n_{1,2}=\pm1$. The part of the integration contour which gives
the dominant contribution is determined by $\rho(u)$ and the relative
sizes of $\omega$ and $m_s$. Let us deal with the two cases
$\omega>m_s$ and $\omega<m_s$ in turn.
\subsubsection{$\omega>m_s$} 
For $n_1=-n_2$, $\rho(u)$ has a stationary point at $a=\omega /
\sqrt{\omega^2-m_s^2}$ and we can apply the method of stationary phase
(see e.g. \cite{1978amms.book.....B}). Since only a small region
around the stationary point contributes to the integral, expanding the
exponent around $a$ gives the leading-order behavior
\begin{align}
I&\sim\f{1}{4}\sqrt{\f{2}{\pi}}\sqrt{m_sR}\f{e^{\rho(a)}}{a^{n-1}\sqrt[4]{a^2-1}^3}\int_{-\delta}^{+\delta} ds\;e^{\f{1}{2}\rho''(a)s^2} \nonumber \\
&\sim\f{1}{2}\Big(\f{\sqrt{\omega^2-m_s^2}}{\omega}\Big)^{n-1}e^{iRn_1\sqrt{\omega^2-m_s^2}}e^{in_1\pi}.
\end{align}
For $n_1=n_2$, $\rho(u)$ no longer has any stationary points in the integration domain, so the
leading order behavior is obtained by integration by parts. Since the
integrand goes to zero at the upper endpoint $+\infty$, the only
contribution will come from the lower endpoint, which as we have
discussed must exactly cancel with the corresponding terms from
\eqref{boundary_term}. The complete leading order behavior of
$C_n(R;m_s,\omega)$ (and similarly $S_n(R;m_s,\omega)$) for
$\omega>m_s$ is hence given by
%
\begin{align}
C_n(R;m_s,\omega)\sim\mathrm{cos}(\omega R)-\Big(\f{\sqrt{\omega^2-m_s^2}}{\omega}\Big)^{n-1}\mathrm{cos}(R\sqrt{\omega^2-m_s^2})\,, \nn\\
S_n(R;m_s,\omega)\sim\mathrm{sin}(\omega R)-\Big(\f{\sqrt{\omega^2-m_s^2}}{\omega}\Big)^{n-1}\mathrm{sin}(R\sqrt{\omega^2-m_s^2})\,.
\end{align}
%
\subsubsection{$\omega<m_s$}
The case $\omega<m_s$ is somewhat more subtle. Now the first
derivative of $\rho(u)$ can only vanish on the imaginary axis. We must
therefore consider the analytic properties of $\rho(u)$ and use the
method of steepest descent \cite{1978amms.book.....B}. The central idea
behind this method is to deform the integration contour in such a way
that it follows lines of constant phase (lines of steepest descent), in
the hope that along the new contour the integral may be evaluated
asymptotically. Firstly, we must take care of the fact that $\rho(u)$
is not analytic in the complex plane, owing to the fact that the square
root term makes it double-valued. Working on a two-sheeted Riemann
surface we can still apply the method of steepest descent, provided the
deformed contour does not include either of the branch points that
appear on the real axis at $u=\pm1$. Since we are extending our
exponent into the complex plane, we must first fix the branch of
$\sqrt{u^2-1}$ that we are using. This will then determine the
constant phase contours and the location of the saddle points. The
asymptotic behavior obtained in the end will of course be independent
of the choice of branch. Writing $u=w+iv$ and $\sqrt{u^2-1}=U+iV$, we
define the principal branch of the square root to be
\begin{align}
\label{principal_branch}
\sqrt{u^2-1}=
\begin{cases}
U+iV & w,v>0 \\
U-iV & w>0,v<0 \\
-U-iV & w,v<0 \\
-U+iV & w<0,v>0
\end{cases}
\end{align}
The second branch of the square root is then simply the negative of
\eqref{principal_branch}.

For $n_1=n_2$, there are no saddle points in the principal Riemann
sheet, so we proceed with a straightforward integration by parts. As
before, since the integrand vanishes at the upper endpoint $+\infty$
the only contribution comes from the lower endpoint, and this will
exactly cancel the corresponding contribution from
\eqref{boundary_term}.

The case $n_1=-n_2$ is somewhat more challenging. In the principal
Riemann sheet, we find two saddle points on the imaginary axis at
$u=b_\pm=\pm i\omega/\sqrt{m_s^2-\omega^2}$.  At these saddle points
the imaginary axis intersects another constant phase contour that
closes on the real axis at $u=\pm m_s/\sqrt{m_s^2-\omega^2}$.

\begin{figure}[htb]
\includegraphics[width=8cm,clip=true]{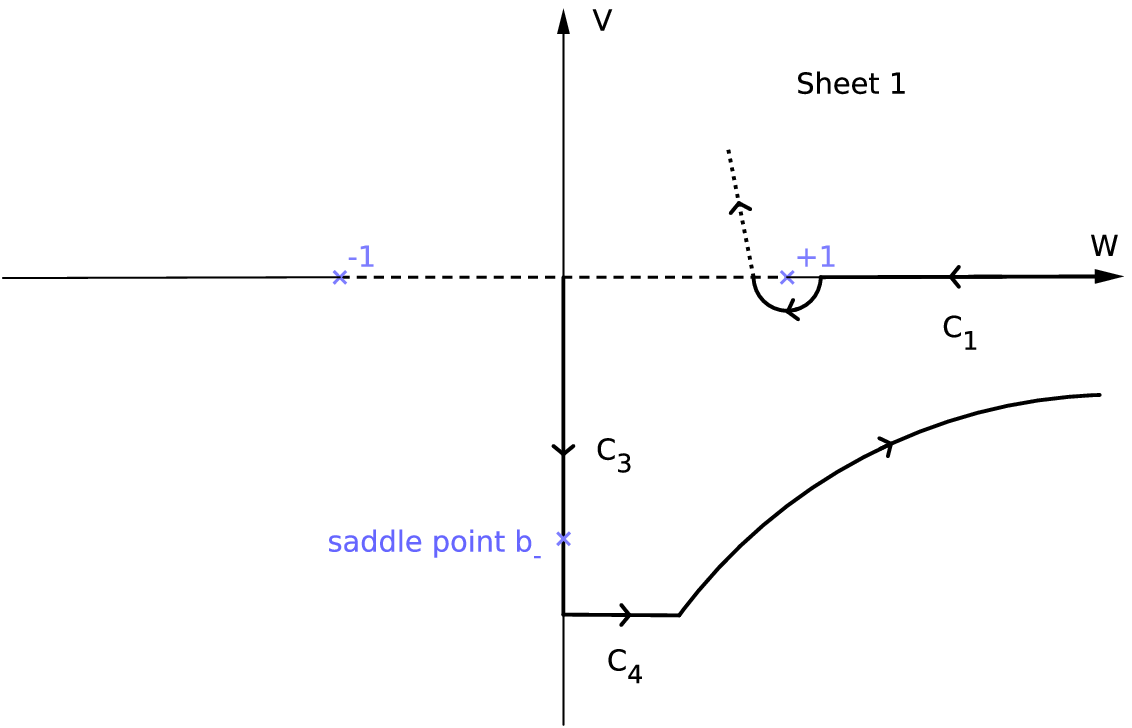}
\includegraphics[width=8cm,clip=true]{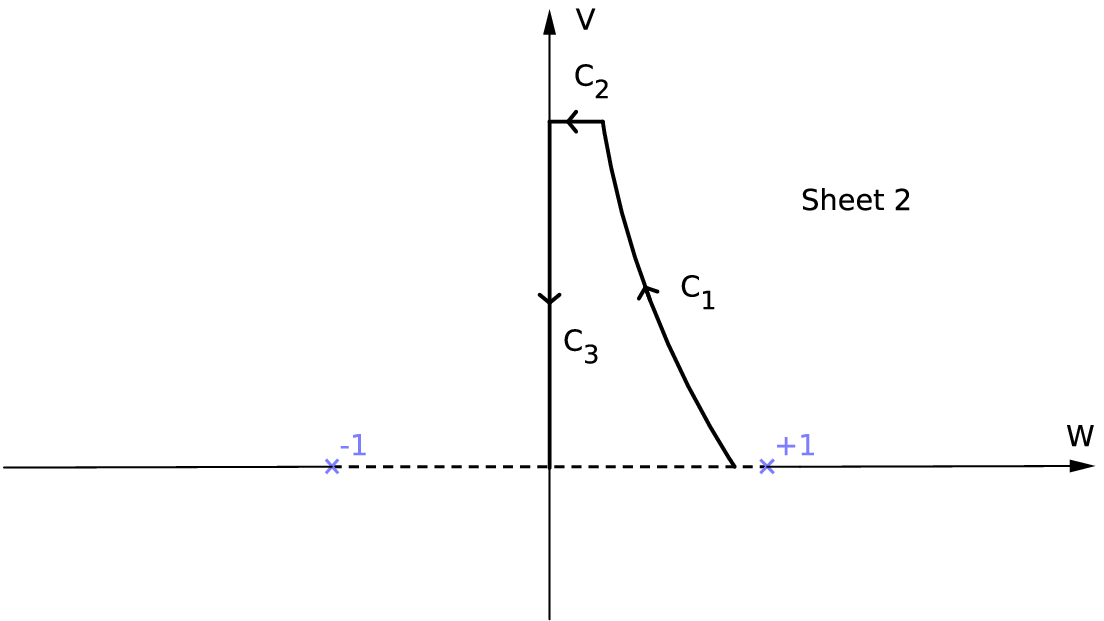}
\caption{Deformed integration contour $C_1+C_2+C_3+C_4$ along lines of
  steepest descent for $\rho(u)=iR(\omega u-m_s\sqrt{u^2-1})+i3\pi/4$
  (corresponding to $n_1=+1,\;n_2=-1$) on the two sheets of the
  Riemann surface associated with $\sqrt{u^2-1}$. Here only the saddle
  point on the negative imaginary axis contributes.}
\label{fig:integration_contour}
\end{figure}

Let us consider the case $n_1=+1$. The original integration contour
runs along the real axis from $1+\epsilon^2/2$ to $+\infty$. We now
deform the contour by going along constant phase lines in the
direction in which the real part of the exponent $\rho(u)$ decreases
(see figure \ref{fig:integration_contour}).  Starting from the lower
endpoint $1+\epsilon^2/2$, we follow the contour $C_1$ with phase
$\omega-m_s\epsilon$ into the lower half of the complex plane, and then
through the branch cut onto the second sheet of the Riemann
surface. On this sheet there are no saddle points, and our constant
phase contour approaches $(\omega-m_s\epsilon)/(m_s+\omega)\pm
i\infty$. We can then connect it to the imaginary axis by a path $C_2$
parametrized by $t+iT$, with $t$ running from
$(\omega-m_s\epsilon)/(m_s+\omega)$ to $0$ and constant $T\gg1$. The
contribution from this path vanishes as $T\rightarrow\infty$. The
integration contour then follows $C_3$ along the imaginary axis
through the branch cut onto the first sheet, through the saddle point
at $b_-=-i\omega/\sqrt{m_s^2-\omega^2}$ and towards $-\infty$. From
there it can be closed onto the positive real axis by the path $C_4$
(analogous to $C_2$) that ultimately makes no contribution. Since the
integrand vanishes as $u\rightarrow+\infty$ the upper endpoint of the
integration is once again unimportant. For the case $n_1=-1$, we can
proceed in a similar fashion, only this time the deformed contour will
pass through the saddle point $b_+$.

We now have all of the ingredients we need to evaluate the
integral. Since our deformed contour together with the original
contour does not include any of the branch points, we can still apply
the Cauchy theorem and approximate the integration along the contour
$C_1+C_2+C_3+C_4$ to obtain the asymptotic behavior of $I$ as
$R\rightarrow\infty$. The two crucial regions of the deformed contour
are the lower endpoint on the contour $C_1$ and the saddle point on the
imaginary axis ($b_-$ for $n_1=+1$ or $b_+$ for $n_1=-1$). Even though
the contribution from the saddle point is sub-dominant, it \emph{does}
give the asymptotic behavior of the original integrals
\eqref{nasty_integrals} that we require; recall again that the
contribution from the lower endpoint will be exactly canceled by the
contribution from \eqref{boundary_term}.  Integrating through the
saddle point $b_\pm$ along the imaginary axis and parameterizing
$u=i(b_\pm+t)$, we obtain
%
%
\begin{align}
\label{i_plus}
I_\pm&\sim\f{1}{4}\sqrt{\f{2}{\pi}}\sqrt{m_sR}\f{e^{\rho(b_\mp)}}{b_\mp^{n-1}\sqrt[4]{b_\mp^2-1}^3}\int_{-\delta}^\delta dt\;ie^{-\f{1}{2}\rho''(b_\mp)t^2} \nonumber \\
&\sim-\f{1}{2}\Big(\mp\f{\sqrt{m_s^2-\omega^2}}{i\omega}\Big)^{n-1}e^{-R\sqrt{m_s^2-\omega^2}}.
\end{align}
where $I_\pm$ corresponds to $n_1=\pm1$. The complete leading-order
behavior of $C_n(R;m_s,\omega)$ (and similarly $S_n(R;m_s,\omega)$)
for $\omega<m_s$ is then given by
\begin{widetext}
\begin{align}
&C_n(R;m_s,\omega)\sim\mathrm{cos}(\omega R)-\Big(\f{\sqrt{m_s^2-\omega^2}}{\omega}\Big)^{n-1}e^{-R\sqrt{m_s^2-\omega^2}}\f{i^{n-1}+(-i)^{n-1}}{2} \nn \\
&S_n(R;m_s,\omega)\sim\mathrm{sin}(\omega R)-\Big(\f{\sqrt{m_s^2-\omega^2}}{\omega}\Big)^{n-1}e^{-R\sqrt{m_s^2-\omega^2}}\f{i^{n-1}-(-i)^{n-1}}{2}.
\end{align}
\end{widetext}

\section{Observational data on compact binaries used in this paper}
\label{sec:binary_appendix}

\begin{center}
\begin{table}[htb]
\caption{Parameters relevant to the binary system PSR J1012+5307 \cite{Lazaridis:2009kq}.}
\label{tab:J1012+5307_parameters}
\begin{tabular}{l r}
\hline
\hline
Period, $P$ (days) & 0.60467271355(3) \\
Period derivative (observed), $\dot{P}^{\rm obs}$ & $5.0(1.4)\;10^{-14}$ \\
Period derivative (intrinsic), $\dot{P}^{\rm intr}$ & $-1.5(1.5)\;10^{-14}$ \\
Mass ratio, q & 10.5(5) \\
NS Mass, $m_1$ ($M_\odot$) & 1.64(22) \\
WD Mass, $m_2$ ($M_\odot$) & 0.16(2) \\
Eccentricity, $e$ ($10^{-6}$) & 1.2(3) \\
\hline
\hline 
\end{tabular}
\end{table}
\end{center}

\subsubsection{PSR J1012+5307}
PSR J1012+5307 is a 5.3ms pulsar in a 14.5hr quasicircular binary
system with a low-mass WD companion \cite{1995MNRAS.273L..68N}. The
relevant parameters for this system are listed in Table
\ref{tab:J1012+5307_parameters}. The parameter values are taken
directly from \cite{Lazaridis:2009kq}. The mass ratio and individual
masses were determined in \cite{1998MNRAS.298..207C}, and the
intrinsic period derivative, corrected for Doppler effects, was
determined in \cite{Lazaridis:2009kq}.

Using the parameters listed in Table \ref{tab:J1012+5307_parameters}
and Eq.~\eqref{pdotgrcirc}, the value of the period derivative
predicted by GR is given by
\begin{align}
\dot{P}_{GR}=-\f{192\pi}{5}\f{q\;m^\f{5}{3}}{(1+q)^2}\Big(\f{2\pi}{P}\Big)^\f{5}{3}=-1.1(2)\times10^{-14}.
\end{align}
Using the method described in section \ref{sec:Pdot_bounds} we obtain
the bound on $\xi$ (and hence $\omega_{\rm BD}$) as a function of the
scalar mass which is displayed in Figure \ref{fig:bounds_all} by a
solid green line. In particular, we find a lower bound $\omega_{\rm
  BD}>1250$ for $m_s<10^{-20}\mathrm{eV}$. The limiting factor here is
our ability to obtain a precise value for the intrinsic period
derivative, once Doppler effects have been accounted for.
\subsubsection{PSR J0751+1807}
PSR J0751+1807 is a millisecond pulsar in a 6hr circular binary system
with a helium WD companion \cite{1995ApJ...453..419L}. The period
derivative has been measured to $\sim 15\%$ precision, after kinematic
corrections have been made. However, the determination of the masses
of the stars in this system has proved to be more of an
issue. Assuming GR to be true, Nice et al. \cite{Nice:2005fi} used
combined observations of the Shapiro delay and orbital period
derivative to constrain the masses of the component stars to a
precision of $\sim 10\%$. Unfortunately, in the context of using the
measured period derivative to constrain modified theories of gravity,
we cannot assume GR in the calculation of the masses. The solution to
this issue is to use only the observations of the Shapiro delay to
constrain the masses, and use these masses in conjunction with the
observed $\dot{P}$ to compare theory with predictions. The problem
with this is that using the Shapiro delay alone provides a very weak
constraint on the masses, with $\sim 100\%$ uncertainty for each of
the two components.  Nonetheless, we could perform an analysis similar
to that done for PSR J1012+5307. Given the large uncertainties
associated with this system, however, we expect that the bounds
obtained from such an analysis would be very weak and would not
provide us with any further insight, and for this reason we have
neglected this system.

\begin{center}
\begin{table}[htb]
\caption{Parameters relevant to the binary system PSR J1141-6545
  \cite{Bhat:2008ck}.}
\label{tab:J1141-6545_parameters}
\begin{tabular}{l r}
\hline
\hline
Period, $P$ (days) & 0.1976509593(1) \\
Period derivative (observed), $\dot{P}^{\rm obs}$ & $-4.03(25)\;10^{-13}$ \\
Period derivative (intrinsic), $\dot{P}^{\rm intr}$ & $-4.01(25)\;10^{-13}$ \\
Mass ratio, q & 1.245(14)\\
NS Mass, $m_1$ ($M_\odot$) & 1.27(1) \\
WD Mass, $m_2$ ($M_\odot$) & 1.02(1) \\
Eccentricity, $e$ & 0.171884(2) \\
Periastron advance, $\dot{\omega}$ ($^\circ yr^{-1}$) & 5.3096(4) \\
\hline
\hline
\end{tabular}
\end{table} 
\end{center}
\subsubsection{PSR J1141-6545}
PSR J1141-6545 is a 394ms pulsar in a moderately eccentric binary
system with a WD companion \cite{2000ApJ...543..321K}. The relevant
parameters for this system are displayed in Table
\ref{tab:J1141-6545_parameters}, and are taken directly from
\cite{Bhat:2008ck}. The masses of the WD and NS were determined by
\cite{Bhat:2008ck}.

This system is comfortably the most useful in the context of putting
bounds on $(\omega_{\rm BD},m_s)$, and in constraining alternative
theories of gravity using observations of the orbital period
derivative in general.  $\dot{P}$ has been measured to remarkable
precision, currently $\sim$6\%, and this is expected to improve
further to $\sim$2\% by 2012 \cite{Bhat:2008ck}. The other necessary
parameters for our purposes, the masses and the periastron shift, have
also been measured to excellent precision, so the total uncertainty in
the system is (relatively) very small. Unfortunately this system does
not have negligible eccentricity, so the result for $\dot{P}$ derived
here does not strictly hold. In order to do a full and accurate
analysis of this system, the result \eqref{period_decay_mbd} must be
generalized to cover eccentric binaries. For the moment we present a
rather crude analysis of this system where we neglect the
eccentricity, to find at least a ball park estimate of the bounds that
we may expect to obtain once a full analysis is performed.  Using the
above parameters and equation \eqref{pdotgrecc}, the value of the
period derivative predicted by GR is given by
\begin{align}
\dot{P}_{GR}=-\f{4q}{(1+q)^2}\f{8}{15\sqrt{3}}\Big(\f{P}{2\pi}\Big)^\f{3}{2}\dot{\omega}^\f{5}{2}=-3.440(3)\;10^{-13}\,.
\end{align}
Using the method described in section \ref{sec:Pdot_bounds}, we obtain
the bound on $\xi$ (and hence $\omega_{\rm BD}$) displayed in Figure
\ref{fig:bounds_all} by a solid blue line. Once we account for
eccentricity in a proper way, this system is very likely to provide
the most stringent bounds among all of the binaries observed so far.

%
%
%

\end{document}